\documentclass[10pt]{article}
\pdfoutput=1
\usepackage[utf8]{inputenc}
\usepackage[T1]{fontenc}
\usepackage[pdftex]{geometry}
\usepackage[english]{babel}
\usepackage[authoryear]{natbib}
\usepackage{textcomp,amsmath,amssymb,pdflscape,longtable,afterpage,calc,booktabs}
\usepackage{url}
\usepackage{graphicx}
\usepackage{float}
\usepackage{multirow}
\usepackage{multicol}
\usepackage{amsfonts,soul,rotating}
\usepackage[version=4]{mhchem}

\newcommand{\drv}[2]{\frac{\mathrm{d}#1}{\mathrm{d}#2}}  

\begin{document}
\let\WriteBookmarks\relax
\def\floatpagepagefraction{1}
\def\textpagefraction{.001}
\thispagestyle{empty}
\begin{center}
\Large Lukas Manske\textsuperscript{1,2}\\[2ex]
Thomas Ruedas\textsuperscript{1,3}\\[2ex]
Ana-Catalina Plesa\textsuperscript{3}\\[2ex]
Philipp Baumeister\textsuperscript{2,3,4}\\[2ex]
Nicola Tosi\textsuperscript{3}\\[2ex]
Natalia Artemieva\textsuperscript{1,5}\\[2ex]
Kai Wünnemann\textsuperscript{1,2}\\[5ex]
\textbf{The influence of interior structure and thermal state on impact melt generation upon large impacts onto terrestrial planets}\\[5ex]
final version\\[5ex]
24 April 2025\\[10ex]
published as:\\
\textit{Journal of Geophysical Research~-- Planets} 130, doi:10.1029/2024JE008481 (2025)\\[15ex]
\normalsize
\textsuperscript{1}Museum für Naturkunde Berlin, Leibniz Institute for Evolution and Biodiversity Science, Germany\\
\textsuperscript{2}Free University of Berlin, Department of Earth Sciences, Germany\\
\textsuperscript{3}Institute of Space Research, German Aerospace Center (DLR), Berlin, Germany\\
\textsuperscript{4}Technische Universität Berlin, Center for Astronomy and Astrophysics, Germany\\
\textsuperscript{5}Planetary Science Institute, Tucson AZ, U.S.A.\\
\rule{0pt}{12pt}
\end{center}
\vfill
This author pre-print version is shared under the Creative Commons Attribution License (CC BY 4.0).
\normalsize
\title{The influence of interior structure and thermal state on impact melt generation upon large impacts onto terrestrial planets}\\[5ex]
\author{Lukas Manske\textsuperscript{1,2}\thanks{Corresponding author: \texttt{lukas.manske@fu-berlin.de}}, Thomas Ruedas\textsuperscript{1,3}, Ana-Catalina Plesa\textsuperscript{3}, Philipp Baumeister\textsuperscript{2,3,4}, Nicola Tosi\textsuperscript{3},\\Natalia Artemieva\textsuperscript{1,5}, Kai Wünnemann\textsuperscript{1,2}}


\date{}
\maketitle

\textbf{Highlights}
\begin{itemize}
\item We quantify impact-induced melt production throughout the thermochemical evolution of planets of various sizes and structures.
\item Apart from shock melting we quantify decompression and plastic work melting, which may contribute significantly to total melt production.
\item We discuss to what extent classical scaling laws are applicable to melt production and derive an empirical scaling law from our results.
\end{itemize}

\begin{abstract} 
We investigate the melt production of planetary impacts as a function of planet size ($R/R_\mathrm{Earth}$=0.1--1.5), impactor size ($L$=1--1000 km), and core size ratio ($R_\mathrm{core}/R$=0.2--0.8) using a combination of parameterized convection models and fully dynamical 2D impact simulations. To this end, we introduce a new method to determine impact-induced melt volumes which we normalize by the impactor volume for better comparability. We find that this normalized melt production, or melting efficiency, is enhanced for large planets when struck by smaller impactors, while for small planets, melting efficiency is elevated when impacted by larger impactors. This diverging behavior can be explained by the thickness of the planets' thermal boundary layer and the shapes of their thermal and lithostatic pressure profiles. We also find that melting efficiency maxima are usually highest on Earth-size planets. We show that the melting efficiency is only affected by core size ratio for large cores and older planets, where melt production is decreased significantly compared to smaller core size ratios. Projecting the lunar impactor flux on the generic planets, we find that Moon-sized planets produce the most melt throughout their evolution, relative to planet volume. Contrary to previous scaling laws, our method accounts for melt production by decompression or plastic work in addition to shock melting. We find that traditional scaling laws underestimate melt production on length scales where variations in the target planets' lithology, temperature, and lithostatic pressure become significant. We propose empirical formulas to predict melt generation as a function of radial structure and thermal age.
\end{abstract}

\section*{Plain Language Summary} 
We use computer simulations to determine how planetary impacts cause melting, considering factors like planet size, impactor size, and core size. We find that larger planets experience more melting when struck by smaller impactors, while on smaller planets more melt is generated with larger impactors. This difference can be explained by the interplay of the temperature distribution inside the planet and its internal pressure, factors that control the amount of external energy needed to cause melting at a given depth. We usually find the largest melt volumes relative to the impactor volume on Earth-sized planets. The planet's core size does not strongly influence melting, except for very old planets with large cores, where melting decreases significantly. In order to estimate the total impact melt volume produced on a planet throughout its evolution, we combine the melt calculation findings with a planet-specific impactor flux function, which describes the size and frequency of impactors striking the planet over a given period of time. A Moon-sized planet accumulates the most melt relative to its size throughout its evolution. We also compare our results to simple melting approximations (``scaling laws''), which fail for melting events in complex planet structures and temperatures. Consequently, we propose a new way of predicting melting based on a planet's structure and age that is better suited for application to real planets than the scaling laws for homogeneous targets and should improve melt volume estimates in geological investigations without the need for an expensive and time-consuming numerical computation.

\section{Introduction}
Asteroid impacts spanning several orders of magnitude in terms of impactor size and delivered energy have been known for decades as one of the most important factors of planetary evolution in the Solar System, especially at its early stages more than 4 billion years ago. The morphological effects of impacts such as crater formation have been studied extensively by numerical modeling as well as in experiments, in the latter case often with a certain emphasis on smaller craters \citep[e.g.,][]{Gree:etal82,OKeAh93,BAIvanov08b,Guld:etal15,Prie:etal17,Kenk:etal18,Alli:etal23b}. Relationships based on dimensional analysis between the impact parameters and geometrical properties of the crater have been established in this way and applied to geological problems with considerable success \citep[e.g.,][]{Croft85,OKeAh93,Holsapple93}. Other consequences of impacts are less conspicuous and more difficult to observe and quantify, for instance the formation of melt, and attempts to relate them to impact parameters have so far mostly been limited to homogeneous targets \citep[e.g.,][]{BjHo87,Pier:etal97}.\par
Finding a relationship between impact parameters and impact melting is complicated by the fact that several processes contribute to the total volume of melt eventually produced. Three mechanisms have to be distinguished in immediate temporal proximity to the impact: 1.~shock melting induced by the almost instantaneous extreme compression and subsequent release of the materials \citep[e.g.,][]{ahrens1972shock,Melo:etal92,osinski2012impact}; 2.~decompression melting caused by the impact-induced uplift of the parts of the target beneath the crater and the associated reduction of the lithostatic pressure  \citep[e.g.,][]{jones2002impact,jones2003impact,jones2005modeling,ivanov2003impacts,Mans:etal21a}; and 3.~plastic work melting due to deformation and internal friction in materials with strength, which has only recently begun to receive due attention \citep{Quin:etal15,KKuGe18,MeIv18,Mans:etal22}. In the longer term, geodynamical processes triggered by the impact may result in further melt production, especially after very large impacts \citep[e.g.,][]{padovan2017impact,JHRoAr-Ha12,Rolf:etal17,RuBr17c,RuBr18b,RuBr19a}. In this paper, we are interested in the direct aftermath of the collision, however, and focus on the first three mechanisms.\par
Another important complication arises in impacts for which important length-scales of the impact process such as the depth of penetration of the impactor approach the thickness of the lithosphere of the target. Early attempts to establish a relation between impact parameters and the total melt volume assumed a homogeneous target \citep[e.g.,][]{BjHo87,ToMe92,ToMe93,Pier:etal97}, but this assumption becomes increasingly invalid as these two types of length-scales approach each other \citep{jones2002impact}. The consequences of a thermally inhomogeneous target structure have first been studied with regard to morphological aspects for example by \citet{Milj:etal13,Milj:etal16} for the Moon, and with a focus on the evolution of the Earth's Hadean crust by \citet{Marc:etal14}. These authors as well as \citet{jones2005modeling} already gave some consideration to the effects of the target's structural heterogeneity on impact-induced melting, but the issue was only recently investigated in more detail and with a more accurate numerical method by \citet{Mans:etal21a} with application to Mars. In particular, the latter authors demonstrated that a homogeneous-target model is insufficient in a specific range of impact magnitudes in which both types of length-scales are similar. However, it performs satisfactorily if either the cold crust/lithosphere or the hot mantle are predominantly affected by the shock. It is worth emphasizing that this range of transitional impact magnitudes is time-dependent: as the lithosphere is thin at the beginning and grows with time, so does the impactor size for which the assumption of a homogeneous target fails. Another recent attempt at a comprehensive representation of impact melt generation is the study by \citet{Naka:etal21}, who express melting-related variables such as the heating of different parts of the mantle due to an impact in terms of Legendre polynomials. This approach in principle allows for a more accurate and geometrically realistic description of the amount and distribution of melt, even though their particular study did not consider the effect of the cold lithosphere and was limited to young planets with a fixed core-to-planet ratio.\par
In this paper, we combine thermal evolution models for a variety of generic planets with impact simulations and investigate the direct aftermath of the collision. We focus on shock, decompression, and plastic work melting mechanisms for different scenarios including different interior structures and thermal states. The growth of the lithosphere that controls the time-dependence of the transitional magnitude range is an aspect of the thermal evolution of the planet. It depends on the planet's internal structure and properties, in particular on its size and its mantle thickness. We generalize the results from \citet{Mans:etal21a} by considering not a single specific planet like Mars but a range of generic terrestrial planets with different sizes and mantle-to-core ratios. In Section \ref{sec:methods} we present our methods including the underlying equations. Our results for different planets are shown in Section \ref{sec:results}, followed by a summary and conclusions in Section \ref{sec:conclusions}.

\section{Methods}\label{sec:methods}
To determine the impact-induced melt production for a variety of generic terrestrial planets, we follow a similar procedure as \citet{Mans:etal21a} in that we consider a single vertical impact into a radially symmetric target. The modeling strategy consists of two steps. First, a parameterized thermal evolution model is used to simulate the global evolution of a target planet with a stagnant lithosphere from the time at which mantle solidification after the accretion phase is completed up to the time of the impact. The radial structure of the target is then used as input for a fully dynamical two-dimensional model of the impact process, which is run until the crater has essentially attained its final form and the production of melt as an immediate consequence of the impact process has ceased; this happens in a matter of hours or days at most, depending on the magnitude of the impact. By generating the pre-impact model with a parameterized one-dimensional evolution model instead of a two- or three-dimensional fully dynamical one as in \citet{Mans:etal21a}, we avoid structural and other complexities introduced by the dynamics and can carry out the impact simulation with a more clearly structured target, which facilitates the analysis of the results. We also make an attempt to derive the material-specific physical properties used in both the evolution and the impact model from the same equation of state for better internal consistency between the models.

\subsection{Planetary thermal and structural evolution modeling}
We consider terrestrial planets of radius $R_\mathrm{P}$, with core mass and radius $M_\mathrm{c}$ and $R_\mathrm{c}$ whose mantles are cooled mostly by convection. Cooling of such a convecting system of thickness $h=R_\mathrm{P}-R_\mathrm{c}$ is driven by the temperature drop $\Delta T$ and controlled by the vigor of convection and the heat flux $q$ through its outer surface. For spherical shells heated from within or below, convective vigor is often characterized by the thermal Rayleigh number, which following \citet{GSchu:etal01} can be written as

\begin{gather}
\mathfrak{Ra}_\mathrm{in}=\frac{4\pi}{3}\left(1-\frac{R_\mathrm{c}}{R_\mathrm{P}}\right)^5\frac{\alpha_\mathrm{m}\varrho^2_\mathrm{m}HG_0R_\mathrm{P}^6}{c_p\kappa^2_\mathrm{m}\eta},\\
\mathfrak{Ra}_\mathrm{b}=\frac{M_\mathrm{c}+\frac{4\pi}{3}\varrho_\mathrm{m} (R_\mathrm{P}^3-R_\mathrm{c}^3)}{\frac{R_\mathrm{P}}{R_\mathrm{c}}-1}\frac{\alpha_\mathrm{m}\Delta T\varrho_\mathrm{m}G_0R_\mathrm{P}}{\kappa_\mathrm{m}\eta}
\end{gather}
for internal and basal heating, respectively, where $\varrho_\mathrm{m}$ is the density, $\eta$ is the viscosity, $c_{p\mathrm{m}}$ is the isobaric specific heat, $\alpha_\mathrm{m}$ is the thermal expansivity, and $\kappa_\mathrm{m}$ is the thermal diffusivity of the convecting layer. The gravity of the planet has been expanded using Newton's law of gravitation with the gravitational constant $G_0$ in $\mathfrak{Ra}_\mathrm{b}$. The Nusselt number, i.e., the ratio of total heat flow and conductive heat flow, for the outer surface of a convecting spherical shell is

\begin{equation}
\mathfrak{Nu}=\frac{q(R_\mathrm{P}-R_\mathrm{c})}{\kappa_\mathrm{m}\varrho_\mathrm{m} c_{p\mathrm{m}}\Delta T}\frac{R_\mathrm{P}}{R_\mathrm{c}}
\end{equation}
\citep[after][]{IwHo97}. The material properties $\varrho_\mathrm{m}$, $\eta$, $c_{p\mathrm{m}}$, $\alpha_\mathrm{m}$, and $\kappa_\mathrm{m}$ as well as the internal heat production $H$ in $\mathfrak{Ra}_\mathrm{in}$ are considered given by the composition of the terrestrial planet, which we assume to be chondritic, and by the mean pressure and temperature of the mantle. The form of $\mathfrak{Ra}$ and $\mathfrak{Nu}$ thus shows that the cooling is strongly controlled by the sizes of the planet and its core, as well as their ratio, and also by $\Delta T$, which itself is also a function of $R_\mathrm{P}$ and $R_\mathrm{c}$ via the radial temperature gradient outside the thermal boundary layers in the mantle. This motivates our decision to test a range of planetary sizes and core-to-planet radius ratios in our generic planets. Specifically, we use radii of 0.1, 0.25, 0.5, 1, and 1.5 times the Earth's radius $R_\mathrm{E}$ and core-to-planet radius ratios of 0.2 (Moon-like), 0.5 (Earth-like), and 0.8 (Mercury-like).\par
The interior structure of the planet is assumed to consist of three layers -- a basaltic crust, a dunitic mantle, and an iron core -- and is constructed by solving a set of three coupled equations that relate the mass $M_\mathrm{P}$ within the radius $r$ of a radially symmetric planet, the density $\varrho$, and the pressure $p$:

\begin{align}
\drv{M_\mathrm{P}(r)}{r}&=4\pi r^2\varrho(r)\\
\drv{p(r)}{r}&=-G_0\frac{M_\mathrm{P}(r)\varrho(r)}{r^2}\\
p(r)&=f\left(\varrho(r),T(r),X(r)\right),
\end{align}
where $f$ is an equation of state that also depends on temperature $T$ and composition $X$. We use the semi-analytical equation of state M-ANEOS \citep{SLThLa72a,Melosh07} for this purpose and integrate down an adiabat with a prescribed initial potential temperature $T_\mathrm{p}$ of 1600 K (or 1400 K in some cases) and interior structure to set the initial condition. The thermal evolution is then simulated by a parameterized one-dimensional model \citep[e.g.,][]{Grot:etal11b,Stam:etal12,Tosi:etal17b,Baum:etal23}; where the algorithm requires constant material properties, we calculate them as averages from the $p$,$T$-dependent ones derived from ANEOS. For the technical details, we refer to Appendix~A of \citet{Baum:etal23}. 

Melting and crustal production are self-consistently calculated during the thermal evolution. We adopt here the approach of \citet{Grot:etal11b} and consider that melting occurs in mantle plumes, assuming that the surface fraction covered by plumes is 0.01 \citep{Grot:etal11b}. The parameters adopted for the thermal evolution models are listed in Tables \ref{tab:evol_param} and \ref{tab:multicol}. We use chondritic abundances for the radioactive heat-producing elements and chondritic solidus and liquidus curves of the form

\begin{equation}
T_\mathrm{s,l}(p)=a_\mathrm{s,l}(p+p_\mathrm{s,l})^{c_\mathrm{s,l}},\label{eq:TslSG}
\end{equation}
established (in slightly different form) by \citet{SiGl29}; $p$ is given in GPa and $T_\mathrm{s,l}$ in K. The construction of the fit and the polynomial approximation used in the thermal evolution models are outlined in \ref{sect:Tslpar}, and the values of the fitting parameters $a_\mathrm{s,l}$, $p_\mathrm{s,l}$, and $c_\mathrm{s,l}$ are given in Table~\ref{tab:evol_param}. This solidus and liquidus are also used for ``dunite'' melting in the impact model, along with a more complicated parameterization of melting for the crustal basalts, also detailed in \ref{sect:Tslpar}. These solidus and liquidus curves are also used in the impact simulations to determine the impact-induced melt in order to ensure consistency and because ANEOS is not capable of providing distinct solidus and liquidus curves.

At 1, 3, and 4.5 Gyr after the formation of the planet, temperature profiles and corresponding crustal thicknesses are saved as input for models of impacts at these times, i.e., the impactors strike radially symmetric targets with lithospheric thicknesses corresponding to these ages. For small planets ($R_\mathrm{P}\leq 0.25 R_\mathrm{E}$) we chose an initial temperature of 1400 K to avoid a higher temperature in the mantle compared to the temperature at the core--mantle boundary (CMB). Due to the overall colder temperature in these cases, no crust was generated during the evolution.

\begin{table}
 \caption{Common model parameters of the thermal evolution models.}
 \centering
 \begin{tabular}{l c}
 \hline
 Surface temperature&288 K\\
 Reference viscosity&10$^{19}$ Pa\,s\\
 Activation energy&325 kJ/mol\\
 Activation volume&1 cm\textsuperscript{3}/mol\\
 Mantle specific heat &1200 J\,kg$^{-1}$\,K$^{-1}$\\
 Crustal thermal conductivity &2 W\,m$^{-1}$\,K$^{-1}$\\
 \multicolumn{2}{l}{\textit{Melting parameterization (Eq.~\ref{eq:TslSG}) for chondritic material}}\\
 Solidus, $p<27.5$ GPa\\
 \hspace{1em}$a_\mathrm{s}$&1337.159 K/GPa\textsuperscript{0.1663}\\
 \hspace{1em}$p_\mathrm{s}$&0.9717 GPa\\
 \hspace{1em}$c_\mathrm{s}$&0.1663\\
 Solidus, $p\geq27.5$ GPa\\
 \hspace{1em}$a_\mathrm{s}$&520.735 K/GPa\textsuperscript{0.4155}\\
 \hspace{1em}$p_\mathrm{s}$&9.6535 GPa\\
 \hspace{1em}$c_\mathrm{s}$&0.4155\\
 Liquidus\\
 \hspace{1em}$a_\mathrm{l}$&310.317 K/GPa\textsuperscript{0.5337}\\
 \hspace{1em}$p_\mathrm{l}$&29.102 GPa\\
 \hspace{1em}$c_\mathrm{l}$&0.5337\\
 \hline
 \end{tabular}
 \label{tab:evol_param}
 \end{table}

\begin{table}[ht]
\caption{Parameters used for the thermal evolution models for an initial potential temperature of 1600\,K.}
\begin{center}
\begin{tabular}{lccccc}
    \hline
    Parameter & $0.1\times R_E$ & $0.25\times R_E$ & $0.5\times R_E$ & $1.0\times R_E$ & $1.5\times R_E$\\\hline
    Planet radius $R_\mathrm{P}$ (km) & 640 & 1600 & 3200 & 6400 & 9600\\[1ex]
    Initial potential temperature $T_\mathrm{pot}$ (K) & 1400 & 1400 & 1600 & 1600 & 1600\\[1ex]
    \multicolumn{3}{c}{\textit{Moon-like CMR} ($R_\mathrm{c}/R_\mathrm{P}=0.2$)}\\
    Core radius $R_c$ (km) &  &  & 640 & 1280 & 1920\\
    Mantle density $\rho_\mathrm{m}$ (kg\,m$^{-3}$) &  &  & 3370.42 & 3996.61 & 4578.43\\
    Core density $\rho_\mathrm{c}$ (kg\,m$^{-3}$) &  &  & 8211.9 & 10954.38 & 12849.05\\
    Surface gravity $g$ (m\,s$^{-2}$) &  &  & 3.05 & 7.25 & 12.47\\
    Core isobaric specific heat $c_{p\mathrm{c}}$ (J\,kg$^{-1}$\,K$^{-1}$) &  &  & 840.0 & 830.82 & 703.14\\
    Temperature at CMB $T_\mathrm{CMB}$ (K) &  &  & 1928.78 & 2585.5 & 2356.77\\
    Mantle thermal expansivity $\alpha_\mathrm{m}$ ($10^{-5}\,\mathrm{K}^{-1}$) &  &  & 2.75 & 1.35 & 0.215\\
    Mantle thermal conductivity $k_\mathrm{m}$ (W\,m$^{-1}$\,K$^{-1}$) &  &  & 3.06 & 8.45 & 23.32\\[1ex]
    \multicolumn{3}{c}{\textit{Earth-like CMR} ($R_\mathrm{c}/R_\mathrm{P}=0.5$)}\\
    Core radius $R_c$ (km) & 320 & 800 & 1600 & 3200 & 4800\\
    Mantle density $\rho_\mathrm{m}$ (kg\,m$^{-3}$) & 3314.26 & 3314.26 & 3388.69 & 4107.25 & 4422.09\\
    Core density $\rho_\mathrm{c}$ (kg\,m$^{-3}$) & 9574.24 & 9574.24 & 8951.1 & 12249.99 & 14437.01\\
    Surface gravity $g$ (m\,s$^{-2}$) & 0.73 & 1.83 & 3.65 & 9.17 & 15.23\\
    Core isobaric specific heat $c_{p\mathrm{c}}$ (J\,kg$^{-1}$\,K$^{-1}$) & 840.0 & 840.0 & 840.0 & 799.09 & 693.0\\
    Temperature at CMB $T_\mathrm{CMB}$ (K) & 1431.9 & 1601.7 & 2210.79 & 4112.15 & 6193.72\\
    Mantle thermal expansivity $\alpha_\mathrm{m}$ ($10^{-5}\,\mathrm{K}^{-1}$) & 2.97 & 2.95 & 2.89 & 2.22 & 1.45\\
    Mantle thermal conductivity $k_\mathrm{m}$ (W\,m$^{-1}$\,K$^{-1}$) & 2.21 & 2.30 & 2.58 & 4.91 & 9.40\\[1ex]
    \multicolumn{3}{c}{\textit{Mercury-like CMR} ($R_\mathrm{c}/R_\mathrm{P}=0.8$)}\\
    Core radius $R_c$ (km) &  &  & 2560 & 5120 & 7680\\
    Mantle density $\rho_\mathrm{m}$ (kg\,m$^{-3}$) &  &  & 3353.42 & 3900.07 & 5112.07\\
    Core density $\rho_\mathrm{c}$ (kg\,m$^{-3}$) &  &  & 9335.18 & 13217.67 & 20154.43\\
    Surface gravity $g$ (m\,s$^{-2}$) &  &  & 5.74 & 15.51 & 34.39\\
    Core isobaric specific heat $c_{p\mathrm{c}}$ (J\,kg$^{-1}$\,K$^{-1}$) &  &  & 840.0 & 753.12 & 693.0\\
    Temperature at CMB $T_\mathrm{CMB}$ (K) &  & & 2078.78 & 3480.97 & 6283.7\\
    Mantle thermal expansivity $\alpha_\mathrm{m}$ ($10^{-5}\,\mathrm{K}^{-1}$) &  &  & 2.91 & 2.35 & 1.21\\
    Mantle thermal conductivity $k_\mathrm{m}$ (W\,m$^{-1}$\,K$^{-1}$) &  &  & 2.5 & 4.37 & 11.56\\
    \hline
\end{tabular}
\end{center}
\label{tab:multicol}
\end{table}

\subsection{Impact modeling}\label{sect:imp-mod}
Using as targets the thermal states and structures of the generic planets described in the previous section, we carried out a systematic numerical modeling study including more than 200 impact simulations with impactor diameters $L$ of 1, 10, 25, 50, 100, 250, 500, or 1000 km. For this purpose we used the iSALE2D shock physics code \citep[e.g.,][]{Amsd:etal80,Melo:etal92,GSColl:etal04,Wunn:etal06} in combination with M-ANEOS \citep{SLThLa72a,Melosh07}, a package to describe the equation of state for the materials involved, i.e., basalt \citep{Pier:etal05} for the crust of the target planet, dunite \citep{Benz:etal89} as a proxy for its mantle as well as the impactor material, and iron \citep{SLThLa72a} as a proxy for the core of the planet. We performed 2D vertical impact simulations with a central gravity setup, a fixed impact velocity $v_\mathrm{imp}$ of 15 km/s and a resolution of 50 cells per projectile radius (CPPR); issues concerning the choice and variability of the impact angle and velocity are discussed in Sect.~\ref{subsect:impang} and \ref{subsect:impvel}, respectively. Furthermore, the so-called ``ROCK'' model \citep{GSColl:etal04} is used as a strength model to describe the elastic-plastic material behavior. In this study we define the strength of damaged rocks by a friction coefficient of $\mu=0.63$ and a cohesion of $Y_0=10\,\mathrm{kPa}$. The von Mises plastic limit $Y_\mathrm{lim}$ describes the yield strength at infinite pressure and is set at $Y_\mathrm{lim}=2.46$ and 3.26 GPa for damaged basalt and dunite, respectively. Furthermore, the material's strength depends on the temperature and may be softened \citep{ohnaka1995shear}; the shear strength drops to zero when the material approaches the solidus temperature. Table \ref{tab:impact_mat_prop} lists further parameters relevant for the impact simulations.

\subsection{Melt quantification and analysis}\label{sect:mlt-qan}
To assess the melt quantities induced upon impact events and analyze the circumstances that lead to melt production, we use a sophisticated Lagrangian tracer method to track the thermodynamic state of the involved material. In this so-called ``peak shock pressure method'', massless Lagrangian tracers are placed initially in each computational cell of the Eulerian grid to represent the material that the cell initially contains. When the simulation is started, the tracers are moved according to the velocity field and record the changes in several physical properties within the computational cells that they pass. The melt production is then determined by reconstructing the thermodynamic path of the material using the peak shock pressure recorded by the tracers \citep[e.g.,][]{Pier:etal95,Wunn:etal08,ArLu05}. When the material experiences the shock, the measured peak shock pressure corresponds to an entropy increase that leads to melting or vaporization upon decompression back to the initial lithostatic pressure $p_0$ if a certain threshold value has been reached.\par
This method allows for an assessment of melting due to the shock. Though shock melting is assumed to be the most important source \citep[e.g.,][]{ahrens1972shock,okeefe1977impact,Pier:etal97,MeIv18}, the other two aforementioned processes, decompression melting and plastic work melting, can contribute significantly to impact-induced melt generation. Recent studies have modified the peak shock pressure method to account also for plastic work and decompression melting \citep[e.g.,][]{Mans:etal21a,Mans:etal22}, which requires the tracer to record the accumulated energy increase by plastic work $\Delta e_\mathrm{plw}$ as well as the initial and final lithostatic pressure $p_0$ and $p_\mathrm{l}$, respectively.\par
\begin{figure}[H]
    \centering
    \includegraphics[scale=0.5]{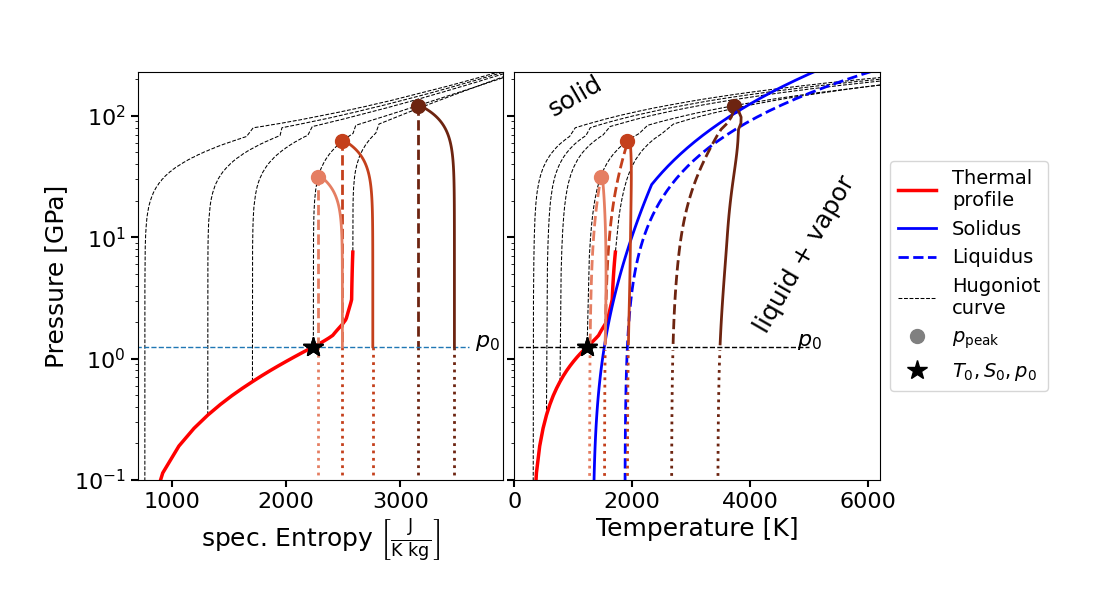}
    \caption{Schematic illustration of the initial thermal profile and a set of Rankine--Hugoniot curves starting from different initial states (black) in a $p$--$S$ (left) and a $p$--$T$ (right) phase diagram. Several thermodynamical paths (orange to brown) describe the shock starting at $p_0$ to the subsequent release from a 30, 60, and 120 GPa peak shock pressure, respectively. The energy increase by plastic work $\Delta e_\mathrm{plw}$ is neglected before reaching the peak shock pressure $p_\mathrm{peak}$. Dashed release paths schematically represent an isentropic release neglecting the effect of plastic deformation.}
    \label{fig:Tz-RK}
\end{figure}

In this study we use the modified method and reconstruct the thermodynamic path of the involved material considering shock, decompression, and plastic work melting in combination with the ANEOS equation of state. Figure~\ref{fig:Tz-RK} illustrates the schematic thermodynamical path reconstructed with this method within a $p$--$S$ and a $p$--$T$ phase diagram, respectively. Based on the initial state of the material (e.g., $z_0$, $T_0$, $p_0$, $S_0$, etc.) we calculate the Rankine--Hugoniot curve, which describes all possible shock states as a function of the peak shock pressure $p_\mathrm{peak}$ that the material experiences. However, one should note that the Rankine--Hugoniot curve does not describe the thermodynamical path, as the shock state is reached instantly upon the shock. When the material reaches the peak shock pressure $p_\mathrm{peak}$, its internal energy has increased substantially. From this peak shock pressure state the material decompresses adiabatically to the final lithostatic pressure in case no plastic work is carried out on it. If the material undergoes plastic deformation, its energy may be increased along the entire thermodynamic path, but most significantly after the shock. At the final state, the impact-induced melt fraction $\psi$ is quantified by comparing the temperature $T_\mathrm{t}$ with the local solidus $T_\mathrm{s}(p_\mathrm{t})$ and liquidus $T_\mathrm{l}(p_\mathrm{t})$:

\begin{align}
\psi&=
\begin{cases}
0,& T_\mathrm{t}\leq T_\mathrm{s}\\
\frac{T_\mathrm{t}-T_\mathrm{s}}{T_\mathrm{l}-T_\mathrm{s}},& T_\mathrm{s}<T_\mathrm{t}<T_\mathrm{l}\\
1,& T_\mathrm{t}\geq T_\mathrm{l}.
\end{cases}
\end{align}
When the melt fraction of the involved material represented by the tracers is determined, the total melt production can be quantified by the melting efficiency, which normalizes the total melt volume by the impactor volume $V_\mathrm{imp}$:

\begin{equation}
    \pi_\mathrm{m}=\frac{\sum\limits_i \psi_i V_i}{V_\mathrm{imp}}.
\end{equation}
Here, $\psi_i$ and $V_i$ are the melt fraction and volume, respectively, of the material represented by the tracer $i$. $V_i$ is defined by the volume of the computational cell the tracer was placed in at the beginning of the simulation.\par

Furthermore, the method allows for distinguishing between shock melting, decompression melting, and plastic work melting. The energy changes induced by these different melting mechanisms can be compared with one another to determine which process dominates in different areas and stages during crater formation. The energy increase by plastic work $\Delta e_\mathrm{plw}$ is recorded directly by the tracers:
\begin{equation}
    \Delta e_\mathrm{plw}= \int_{0}^t \frac{s_{ij}}{\rho}\dot{\varepsilon}_{ij} \;\mathrm{d}t .
\end{equation}
Here, $\dot{\varepsilon}_{ij}$ is the strain rate, $s_{ij}$ describes the deviatoric stress tensor and $\rho$ is the
material's density. A more detailed description can be found in \citet{Mans:etal21a,Mans:etal22}. The energy gain due to the shock $\Delta e_\mathrm{shock}$ can be calculated from the final energy when reconstructing the thermodynamical path as described above. It is defined by the energy difference between the initial energy $e_0$ and the energy after shock and decompression of the material to the initial lithostatic pressure $e_t(p_0)$ while not considering the effect of plastic work ($\Delta e_\mathrm{plw}=0$, isentropic release):

\begin{equation}
    \Delta e_\mathrm{shock}=\left. e_\mathrm{t}(p_0)\right|_{\Delta e_\mathrm{plw}=0}-e_0.
\end{equation}
The energy difference due to decompression melting is calculated via the difference in the average energies of the solidus $e_\mathrm{s}$ and liquidus $e_\mathrm{l}$ at final $p_t$ and initial pressure $p_0$:

\begin{align}
\Delta e_\mathrm{decom}&=
\begin{cases}
\frac{e_\mathrm{l}(p_0)+e_\mathrm{s}(p_0)}{2}-\frac{e_\mathrm{l}(p_\mathrm{t})+e_\mathrm{s}(p_\mathrm{t})}{2}&\text{if $p_\mathrm{t}<p_0$,}\\ 
0&\text{else.}
\end{cases}
\end{align}
In order to assess which melt mechanism contributes to what extent to the total energy difference $\Delta e^\mathrm{tot}_i=\Delta e^\mathrm{shock}_i+\Delta e^\mathrm{plw}_i+\Delta e^\mathrm{decom}_i=\sum_i \Delta e_i^\mu$, we compare the individual contributions with $\Delta e^\mathrm{tot}_i$ measured by each tracer $i$:

\begin{equation}
    \gamma^\mu_i=\frac{\Delta e^\mu_i}{\Delta e^\mathrm{tot}_i}.
\end{equation}
Here, $\mu$ indicates one of the three different processes that add to the total energy difference $\Delta e^\mathrm{tot}$. This allows us to quantify the relative volumetric proportion of melting caused by a given melting mechanism $\mu$ as

\begin{equation}
    m^\mu=\frac{\sum\limits_i \psi_i V_i\gamma_i^\mu}{\sum\limits_i \psi_i V_i}.
\end{equation}

We do not distinguish between impact-induced melting and vaporization. Thus, the presented melt volumes include the volume of vapor, which e.g. in case of dunite at $v_\mathrm{imp} = 15$ km/s is less than 6\% of the total melt and vapor mass \citep{SVETSOV201650}. Furthermore, we only determine the melt of the mantle and crust, but not of the iron core. Due to the hot thermal profiles, low-degree melting occurs in the far field. This melt, however, may be hard to resolve since it requires large simulation domains and may be easily produced by numerical artefacts for instance in the pressure field. This effect is further enhanced by the absence of melting enthalpy (latent heat) in our models due to limitations in ANEOS that do not allow for the implementation of the solidus and liquidus function used in this study. Thus we truncated the entire melt domain at a fixed radius of 5 impactor diameters $L$ around the depth of one impactor diameter and in case of cutting the thermal profiles, we cut 25 K below the solidus instead directly at $T_\mathrm{s}$. Furthermore, we truncate the melt calculation at 2\% melt fraction similar to other studies \citep[e.g.,][]{ToMe93}. As a consequence, the amount of melt is likely an overestimate, and the aforementioned truncations should help to compensate for it to some extent. Moreover, antipodal melt generation upon very large impacts cannot be accounted for, resulting in underestimating melt in such specific cases.

Traditionally, the melt produced by the impact would be plotted against the impactor diameter, which is the only relevant length-scale in the established theoretical models of impact melt generation. In models of a thermally evolving planet, however, the target is not isothermal and homogeneous but has at least a radial thermal and compositional structure that results from the thermal evolution and crust formation of the planet before the impact. This preexisting structure introduces the thicknesses of the thermal lithosphere and the crust as additional, target-specific length-scales that are independent from the impact itself but nonetheless relevant for melt production in an impact. Our models suggest that the thermal structure tends to dominate, i.e., the amount of melt formed by the impact is controlled primarily by the fraction of impactor energy expended on heating a given volume of the target material to the solidus. Hence, we attempt to identify a length-scale that corresponds to the depth where we approximately expect most of the impact-generated melt. That length-scale is also related to the impactor size via the penetration depth on the one hand and to the thermal structure as given by the target geotherm and solidus on the other hand. To this end we calculate a certain depth $d$ along the vertical profile underneath the point of impact where the measured shock pressure exceeds the critical melting pressure necessary to induce melting. Down to this depth the material melts due to the shock, whereby the degree of melting reaches its maximum around $d/2$ below the point of impact and decreases approximately concentrically with increasing distance from this center of melting. We use this center of melting, whose depth we refer to as the depth of melting $d_\mathrm{m}=d/2$, to plot melting efficiency data, since $d_\mathrm{m}$ can be easily compared with other critical depth measures related to the thermal profile, as discussed in the following section. Consequently, $d_\mathrm{m}$ gives a depth estimate at which the largest melt quantities are produced as a function of the impactor size $L$, the planet's thermal and pressure profile as well as other factors such as the material composition and impactor speed. However, the depth of melting should be interpreted as a first-order estimate, since decompression and plastic work melting most likely exceed this depth range. Due to the additional complexity these melting mechanisms would add (cf. Fig.~\ref{fig:Melt_Example}), we stick with the definition described above.

\begin{table}[ht]
\caption{Material parameters used for the impact models.}
\begin{center}
\begin{tabular}{lcc}
    \hline
    Parameter & Dunite & Basalt \\\hline
    Low pressure density (kg m$^{-3}$)&3320&2860\\
    Cohesion (damage) $Y_0$ (kPa)& 10 & 10 \\
    Friction coefficient (damage) $\mu$ & 0.63 & 0.63 \\
    Strength limit (damage) $Y_\mathrm{lim}$ (GPa) & 2.46 & 3.26 \\
    Specific internal energy of melting $E_M$ (J kg$^{-1}$) & $7.1 \times 10^6$ & $4.7 \times 10^6$ \\
    \hline
\end{tabular}
\end{center}
\label{tab:impact_mat_prop}
\end{table}

\section{Results}\label{sec:results}
\subsection{Thermal profiles of generic planets}
\begin{sidewaysfigure}
    \centering
    \includegraphics[scale=1.0]{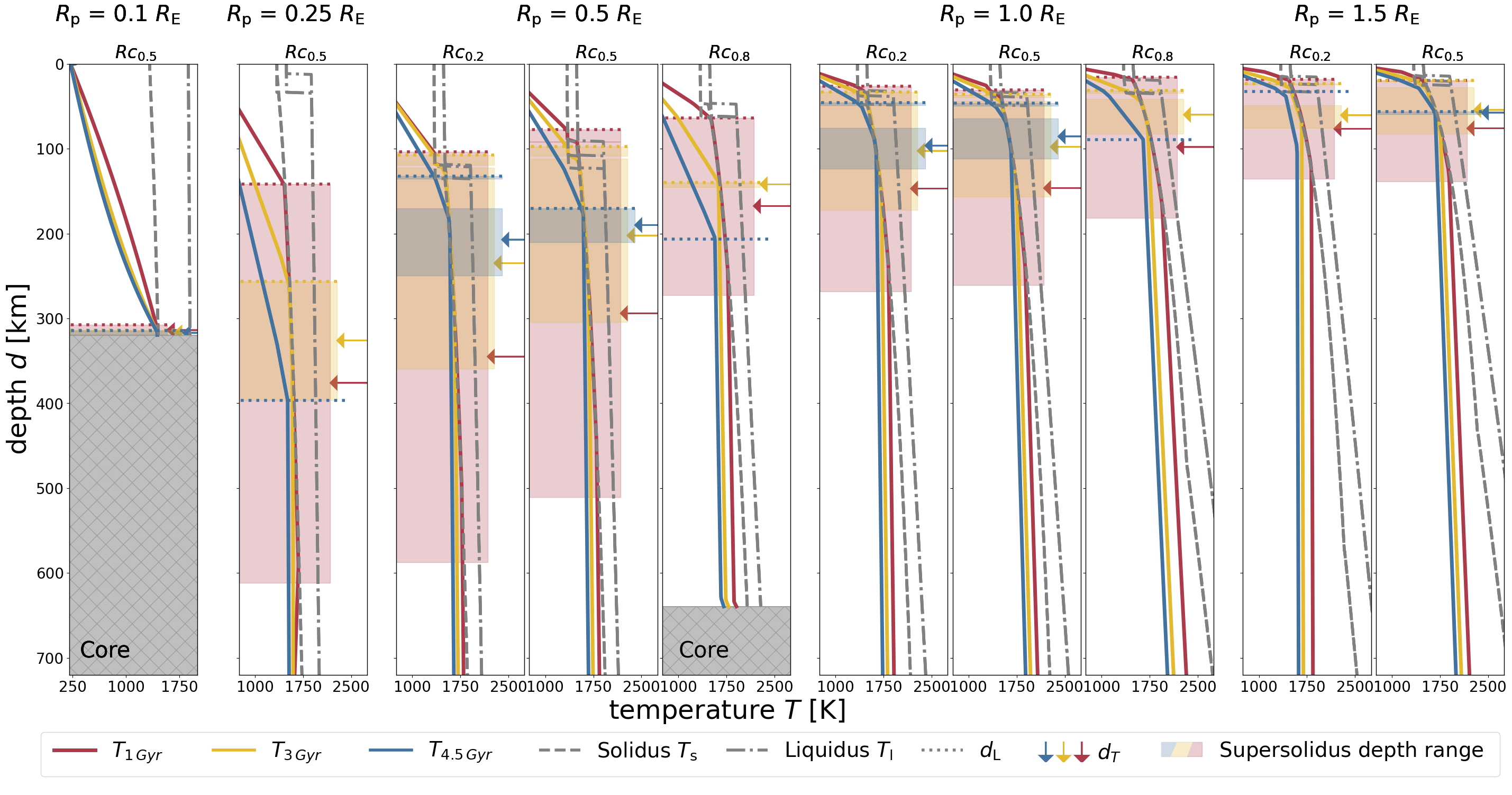}
    \caption{Evolution of the thermal profiles (1--4.5 Gyr) and crust formation for the generic terrestrial planets in the upper $720$~km for better comparability. Dotted horizontal lines mark the thickness of the lithosphere, $d_\mathrm{L}$, the color-filled boxes indicate the initial supersolidus depth ranges, and the colored arrows their centers, $d_\mathrm{T}$. The colors correspond to the ages of the profiles, depths, or depth ranges.}
    \label{fig:Tz}
\end{sidewaysfigure}

Figure \ref{fig:Tz} illustrates the resulting thermal profiles that are used as input for the target planets in the impact models. These profiles describe the state of the thermal evolution at 1, 3, and 4.5 Gyr after planet formation. For all three profiles, we additionally depict the thickness of the lithosphere, denoted as $d_\mathrm{L}$ (marked by colored vertical dotted lines), as well as the depth $d_\mathrm{T}$, indicating the center of the depth range where the temperatures have been reduced to the solidus (indicated by arrows). The latter depth range, if it exists, is marked by the colored area of the same color as the corresponding thermal profile. If it does not exist, we set $d_\mathrm{T}$ equal to $d_\mathrm{L}$ (cf. Sect.\ref{subsect:melteffparam}). In the following, the centered depth of this supersolidus depth range $d_\mathrm{T}$ is referred to as the ``supersolidus depth''. $d_\mathrm{L}$ is calculated by determining the upper boundary of the region where $T \approx T_\mathrm{s}$, if present, or otherwise by the upper thermal boundary layer.

\subsection{Impact-induced melt production on generic planets}
We investigate the effect of the planet's size, structure and thermal evolution on impact-induced melt production. To this end, we carried out a systematic numerical modeling study with more than 200 impact simulations. In particular, we determine and compare the melting efficiency $\pi_\mathrm{m}$ and the effect of the different melting mechanisms $\mu$ of the individual impact scenarios.

\begin{figure}[h!]
    \centering
    \includegraphics[scale=0.3]{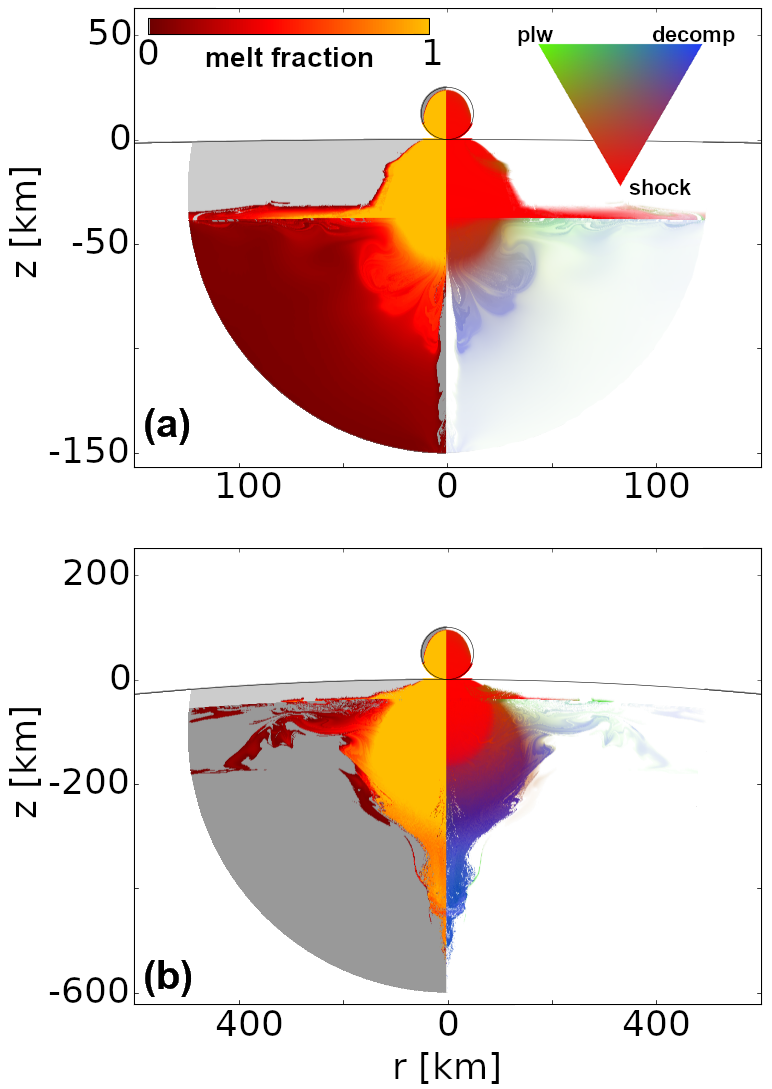}
    \caption{Impact on a 3 Gyr-old Earth-sized planet with a small core $R_\mathrm{c}=0.2R$ and $L=25$~km (a) and $100$~km (b). In each panel, the left half displays the melt fraction and the right half the melting mechanism, with the materials shown at their pre-impact positions. Basaltic crust is shown in light gray and dunitic mantle material in medium-shaded gray. The melt generation farther than $5L$ from the center of the melt region below the impact point (approximated by $z=L$) is neglected.}
    \label{fig:Melt_Example}
\end{figure}

Figure \ref{fig:Melt_Example} illustrates impact-induced melting determined by the method described above with an impact onto a 3 Gyr-old Earth-sized planet with a small core ($R_\mathrm{c}=0.2R_\mathrm{P}$) and impactor diameters $L=25$~km (a) and $100$~km (b) as an example. Here, the material's post-impact thermal state is plotted at its pre-impact position. In the left half of each panel, the impact-induced melt fraction $\psi_i$ is plotted. In the right half the different colors indicate the contribution of melting mechanism $\gamma^{\mu}_{i}$ while the opacity indicates the melt fraction once more. The blank area beyond 5 $L$ from the impact center corresponds to the area where melt production is not tracked anymore as explained in Section~\ref{sect:mlt-qan}. Typically, the highest melt fraction and superheated material originate from the projectile and an approximately spherical volume spreading from the region below the impact point at the depth of melting, where shock pressures are highest and almost constant. This region is often referred to as the ``isobaric core'' \citep{Croft82,Pier:etal95} and is dominated by shock melting, as is the impactor material. 
 Beyond the isobaric core, a combination of all melting mechanisms contributes to melting. Here, the influence of shock melting concentrically decreases, whereby the dominance in the relative proportions of decompression melting and plastic work melting shifts from the former to the latter with decreasing depth. Thus, with increasing distance from the isobaric core, plastic work melting and decompression melting can, under certain conditions, significantly add to melt production \citep[e.g.,][]{Mans:etal21a,Mans:etal22}. If these conditions for the different melting mechanisms $\mu$ are ideal, maxima in melting efficiency $\pi_\mathrm{m}$ are reached, e.g., (1) if the surrounding material is hotter so that critical shock pressures sufficient for generating melt are significantly reduced \citep{Mans:etal21a}; (2) if already hot deep-seated material is subjected to a large pressure drop during uplift in the course of crater formation, resulting in decompression melting; (3) if near-surface material with great strength (low temperatures and pressures) at the edge of the isobaric core is strongly deformed during the cratering process \citep{Mans:etal22}; and (4) if the involved material melts at lower critical shock pressures, such as basalt compared to dunite \citep{stoffler2018shock}. In general, melting efficiency maxima occur at impactor sizes for which the depth where most of the impactor's energy is deposited approaches the depth where the planet's thermal profile is closest to the solidus temperature \citep{Mans:etal21a}. In Figure \ref{fig:Melt_Example}, the pattern of melting in the region beyond the isobaric core is typical for impacts on warm planetary targets. In cold targets, melting typically does not exceed the boundary of the isobaric core or a slightly larger area concentrically spread around it \citep[e.g.,][]{Mans:etal22}.

\subsubsection{Melt production as a function of thermal evolution and melting mechanism}
\begin{figure}
    \centering
    \includegraphics[scale=0.65]{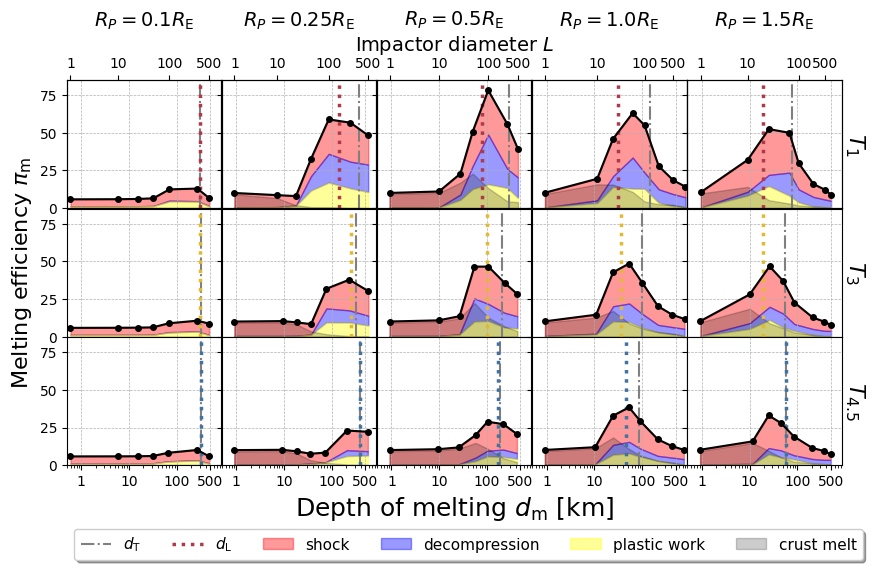}
    \caption{Melting efficiency $\pi_\mathrm{m}$ along with the contribution of the melting mechanisms $\mu$ and melt composition are plotted for different planet ages and planet sizes against the depth of melting $d_\mathrm{m}$ for a medium-sized core ($R_\mathrm{c} = 0.5R_\mathrm{P}$). The depth of the lithosphere $d_\mathrm{L}$ and the supersolidus depth $d_\mathrm{T}$ are plotted according to the thermal profile (Fig.~\ref{fig:Tz}). Melt that does not consist of crustal material is referred to as molten mantle material. For 1 Gyr old planets, the data is also depicted as a function of impactor diameter $L$ for better comparison. The introduced impactor diameter axis also applies approximately to the subsequent rows.}
    \label{fig:Melt_Age}
\end{figure}

Figure \ref{fig:Melt_Age} illustrates the melting efficiency $\pi_\mathrm{m}$ for various impactor and planet sizes at different ages throughout their evolution for a relative core size of $R_\mathrm{c} = 0.5R_\mathrm{P}$. $\pi_\mathrm{m}$ is plotted versus the depth of melting $d_\mathrm{m}$, which correlates with the impactor diameter $L$  (Sect.~\ref{sect:imp-mod}). For planets with radii smaller than $R_\mathrm{E}$, the maximum impactor size was limited to $L\leq 500$ km. Furthermore, the volumetric proportion of melting caused by the different melting mechanisms $m^\mu$ and the fraction of crustal melt are also shown. Additionally, the thickness of the lithosphere $d_\mathrm{L}$ and the supersolidus depth $d_\mathrm{T}$ are indicated by vertical lines.For comparison, the impactor diameter $L$ is plotted for 1 Gyr old planets; this axis also applies approximately to the subsequent rows. In the following figures however, $L$ is not depicted anymore as it does not map uniformly to a given $d_\mathrm{m}$ and thermal profile.

As secular cooling of the planet's interior progresses, the amount of melt produced by impacts decreases substantially. When comparing the peak melt production at 1 and 4.5 Gyr, we observe a reduction in melting efficiency by about 40\% for large ($R_\mathrm{P} = R_\mathrm{E}$ and $1.5 R_\mathrm{E}$) to 60\% for small ($R_\mathrm{P} = 0.25 R_\mathrm{E}$ and $0.5 R_\mathrm{E}$) planets. Peaks in melting efficiency appear at impactor sizes for which the depth of melting is similar to the thickness of the lithosphere, i.e., $d_\mathrm{m} \approx d_\mathrm{L}$. As $d_\mathrm{m}$ is defined to indicate the depth where most of the shock melt is produced for a given impactor size $L$ and the planet's thermal conditions, one may expect that the peak of melting efficiency occurs where the melting depth is similar to the supersolidus depth, $d_\mathrm{m}\approx d_\mathrm{T}$, as the material in this depth requires the lowest critical peak shock pressures for incipient melting. However, the calculation of $d_\mathrm{m}$ is based on shock melting only, whereas peaks in melting efficiency occur in impacts in which the different melting mechanisms $\mu$ combine most productively. We address this topic in more detail in the discussion section below. Besides studying only shock melting, which is the dominant melting mechanism at $v_\mathrm{i} \geq 15$ km/s \citep[cf.][]{Mans:etal22}, we analysed the effects of decompression and plastic work melting in some more detail.

\begin{figure}[h!]
    \centering
    \includegraphics[scale=0.65]{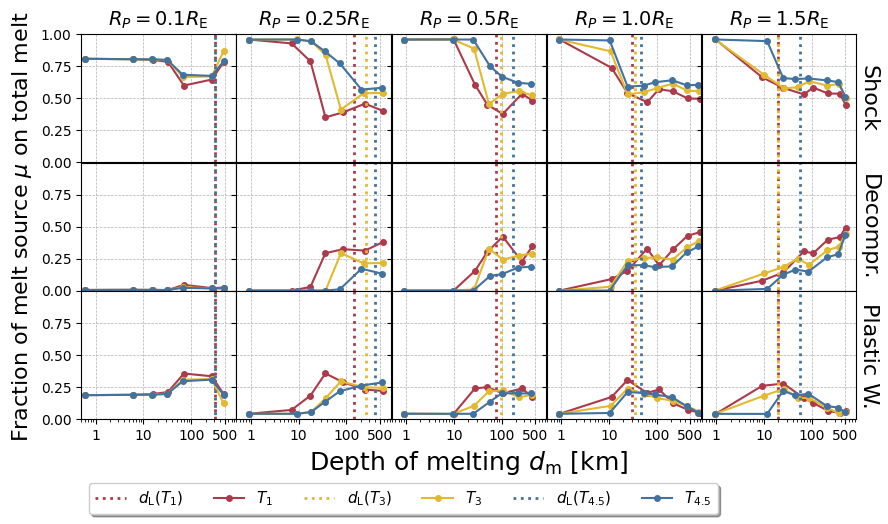}
    \caption{The fraction of different melting mechanisms $\mu$ (top: shock, center: decompression, bottom: plastic work) on total melting is shown as a function of the depth of melting $d_\mathrm{m}$ for different planet sizes and ages and a medium-sized core $R_\mathrm{c} = 0.5 R_\mathrm{P}$. Additionally, the depth of the lithosphere $d_\mathrm{L}$ is plotted according to the thermal profile.}
    \label{fig:Melt_source}
\end{figure}
 Figure \ref{fig:Melt_source} illustrates the contribution of the different melting mechanisms $m^{\mu}$ to total melting versus $d_\mathrm{m}$.  As the impactor size $L$ and thus $d_\mathrm{m}$ increases, the contribution of decompression and plastic work melting to overall melt production becomes significant (cf. right column). Shock melting is dominant for small impactor sizes, whereas the other two melting mechanisms begin to contribute to overall melt production when the depth of melting approaches the thickness of the lithosphere at melting depths of roughly 10--30 km ($L$ = 10--50 km) for young and large and for old and small planets, respectively.\par
The efficiency of plastic work melting depends on the deformation of strong material and thus the dynamics of the crater formation process. The strongest material can typically be found near the surface, since deeper-seated material is thermally weakened. However, the proportion of plastic work melting in the total melt increases when the depth of melting approaches the depth of the lithosphere. Here, the largest contribution of plastic work to total melting lies in the range of 20--35\% among the different scenarios. For $d_\mathrm{m}\lesssim d_\mathrm{L}$, the fraction of plastic work melt is further enhanced on younger planets (1 Gyr) compared to older ones. For $d_\mathrm{m}\gtrsim d_\mathrm{L}$, the plastic work melt contribution stays constant or decreases slowly, especially for large planets where the effect of decompression melting increases. In this range, plastic work melting is slightly enhanced on older planets (4.5 Gyr).\par
The opposed effects of the planet's age on enhanced plastic work melting for small and large depths of melting may be explained by two factors: on the one hand by thermal softening, and on the other hand by the thicker crusts for older planets (4.5 Gyr). For a large depth of melting and thus large impactors, the contribution from crustal melt is negligible. In this case, melt production is reduced in young planets since the strength of the material is more efficiently reduced by thermal softening. For small melting depths and thus small impactors, the thickness of the crust matters, in particular in old planets where it may become relatively large. In that case, the melt production from our evolution models shows that melting is shallow and the size of the crustal contribution to it depends on the crustal thickness and thus the planet's age. We found that material properties of the crust (basalt) strongly favor shock melting over plastic work melting, due to the relatively low critical shock pressure for melting of basalt compared to dunite. For example, plastic work melting is similar on bodies with $R_\mathrm{P} = 0.1 R_\mathrm{E}$ without crust and on larger planets with crust. Thus, plastic work melting significantly enhances melt production if more and more mantle melt is produced
and therefore plastic work melting is reduced in older planets due to their thicker crusts. We note that this behavior may change when using different material strength parameters.\par
Decompression melting, however, is always enhanced on hotter planets (1 Gyr) compared to colder ones (3 and 4.5 Gyr), because the higher mantle temperatures increase the likelihood of melting the material upon upwelling and thus decompression. Its contribution to overall melt production begins beyond a certain critical threshold depth of melting and tends to increase with increasing $d_\mathrm{m}$ up to 50\% for large planets.

\subsubsection{Melt production as a function of planet and core size}
\begin{figure}[h!]
    \centering
    \includegraphics[width=\textwidth]{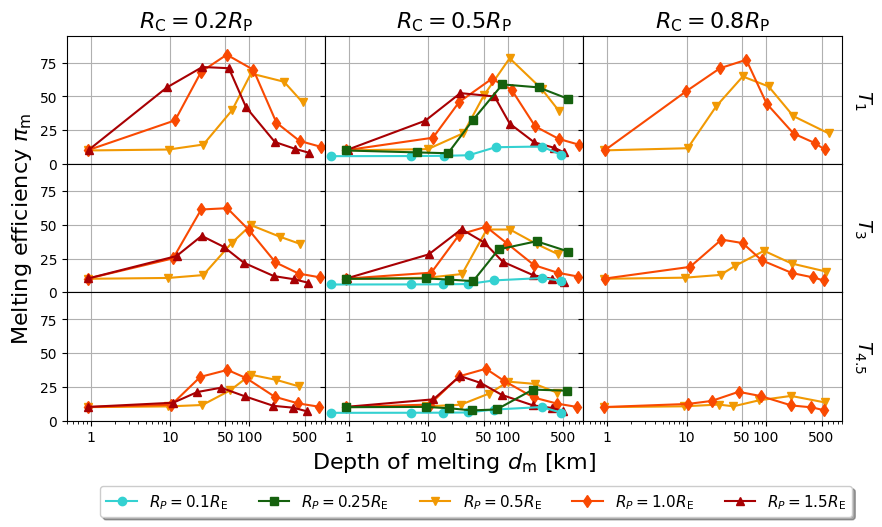}
    \caption{Melting efficiency $\pi_\mathrm{m}$ as a function of the depth of melting $d_\mathrm{m}$ for different planet sizes $R_\mathrm{P}$ for all simulations.}
    \label{fig:melt-pl-size}
\end{figure}

In this section we assess the effects of the size of the planet and its core on impact-induced melting. Figure \ref{fig:melt-pl-size} illustrates the melting efficiency $\pi_\mathrm{m}$ as a function of the depth of melting $d_\mathrm{m}$ for different planet sizes $R_\mathrm{P}$ in all simulations. 

We find that melting efficiency is usually highest in Earth-sized planets, where maxima occur at $d_\mathrm{m} \approx 50$ km. For smaller $d_\mathrm{m}$, we find most efficient melting may shift towards larger planets ($R_\mathrm{P} > R_\mathrm{E}$), whereas for larger $d_\mathrm{m}$, most efficient melting can be found in smaller planets ($R_\mathrm{P} < R_\mathrm{E}$). Thus, our data show the following for sufficiently large $d_\mathrm{m}$, where the impactor penetrates through the upper, relatively cold surface layers: melt production is more efficient on large planets when struck by smaller impactors, while on small planets, it is more efficient with larger impactors. Consequently, peaks in melting efficiency occur at smaller $d_\mathrm{m}$ on larger planets than on smaller ones. The likely reason is the relatively small thickness of the larger planet's lithosphere $d_\mathrm{L}$ compared to small planets of the same age (cf. Fig.~\ref{fig:Tz}). Due to such thin thermal boundary layers on larger planets, smaller impactors and thus smaller depths of melting $d_\mathrm{m}$ match the thickness of the lithosphere $d_\mathrm{L}$ and the supersolidus depth $d_\mathrm{T}$, which results in melting efficiency maxima. For smaller planets the opposite is the case. This result highlights the importance of the interplay between the thermal profile and the solidus function, which is controlled by the pressure gradient and the material. Although the mantle temperatures in larger planets are higher in our data, the solidus $T_\mathrm{s}$ and liquidus $T_\mathrm{l}$ increase significantly with depth due to their comparatively greater lithostatic pressure, leading to a large temperature difference between the geotherm $T(z)$ and the solidus $T_\mathrm{s}$ in the deeper parts of the mantle. Smaller planets, by contrast, have a thicker thermal boundary layer at a given time but a smaller increase in lithostatic pressure with depth, resulting in a massive, deep-seated layer where $T(z)$ is close to $T_\mathrm{s}$. The minimum melt production, however, is independent from the thermal evolution of the planets. It corresponds to the melting efficiency of an impact dominated by shock melting onto a cold target with about $\pi_\mathrm{m}^\mathrm{c} \approx 10.1$ for basalt (crust) and $\pi_\mathrm{m}^\mathrm{m} \approx 5.81$ for dunite (mantle).

\begin{figure}[H]
    \centering
    \includegraphics[width=\textwidth]{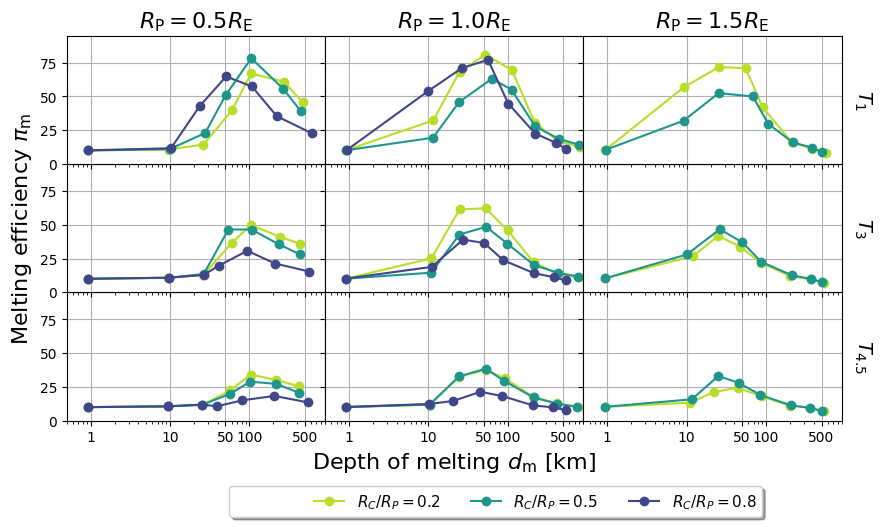}
    \caption{Melting efficiency $\pi_\mathrm{m}$ as a function of the depth of melting $d_\mathrm{m}$ for different core size ratios $R_\mathrm{c}/R_\mathrm{P}$.}
    \label{fig:melt-c-size}
\end{figure}

Figure \ref{fig:melt-c-size} illustrates the melting efficiency $\pi_\mathrm{m}$ as a function of the depth of melting $d_\mathrm{m}$ for different core size ratios $R_\mathrm{c}/R_\mathrm{P}$. In general, the melting efficiency is not strongly affected by the core size ratio except for very large cores ($R_\mathrm{c}/R_\mathrm{P} = 0.8$) and older planets (3 and 4.5 Gyr). In those cases, melt production is decreased significantly compared to smaller core size ratios, which can be explained by the more efficient cooling of the planets with large cores and the reduced fraction of the planet that is endowed with radioactive heat sources. However, since we do not account for the generation of melt in core material, our calculated melt production may be artificially reduced in models in which the depth of the melt region exceeds the depth of the core--mantle boundary, i.e., $2d_\mathrm{m}\gtrsim R_\mathrm{P}-R_\mathrm{c}$. This is due to the fact that in such cases impactors with the corresponding size may transfer a major fraction of their energy into the core rather than the mantle or crust. On the other hand, this apparent reduction in mantle melt production might partly be compensated by double shocks due to reflection of the direct shock wave at the core--mantle boundary.

\section{Discussion}
We have analysed the effects of impact-induced melt production in a range of generic planets, each characterized by a distinct planet and core size as well as the evolutionary state, which defines the thermal profile and the crustal thickness. Our study reveals a tripartite division of melt production, encompassing three distinct melting efficiency regimes: (a) The crust melting regime, where $d_\mathrm{m}$ is situated within the relatively cold crust or lithosphere ($d_\mathrm{m} \ll d_\mathrm{L}$); (b) The peak melting efficiency regime, characterized by $d_\mathrm{m}$ reaching down to the transitional region between the relatively cool thermal boundary layer and the hot convecting deeper interior where the difference between the geotherm and the solidus is smallest ($d_\mathrm{m} \approx d_\mathrm{L}$); (c) The mantle melting regime, with $d_\mathrm{m} \gg d_\mathrm{L}$, located in the deep subsolidus mantle.

\label{Discussion}
\subsection{Melting efficiency parameterization}
\label{subsect:melteffparam}
\begin{figure}[H]
    \centering
    \includegraphics[width=\textwidth]{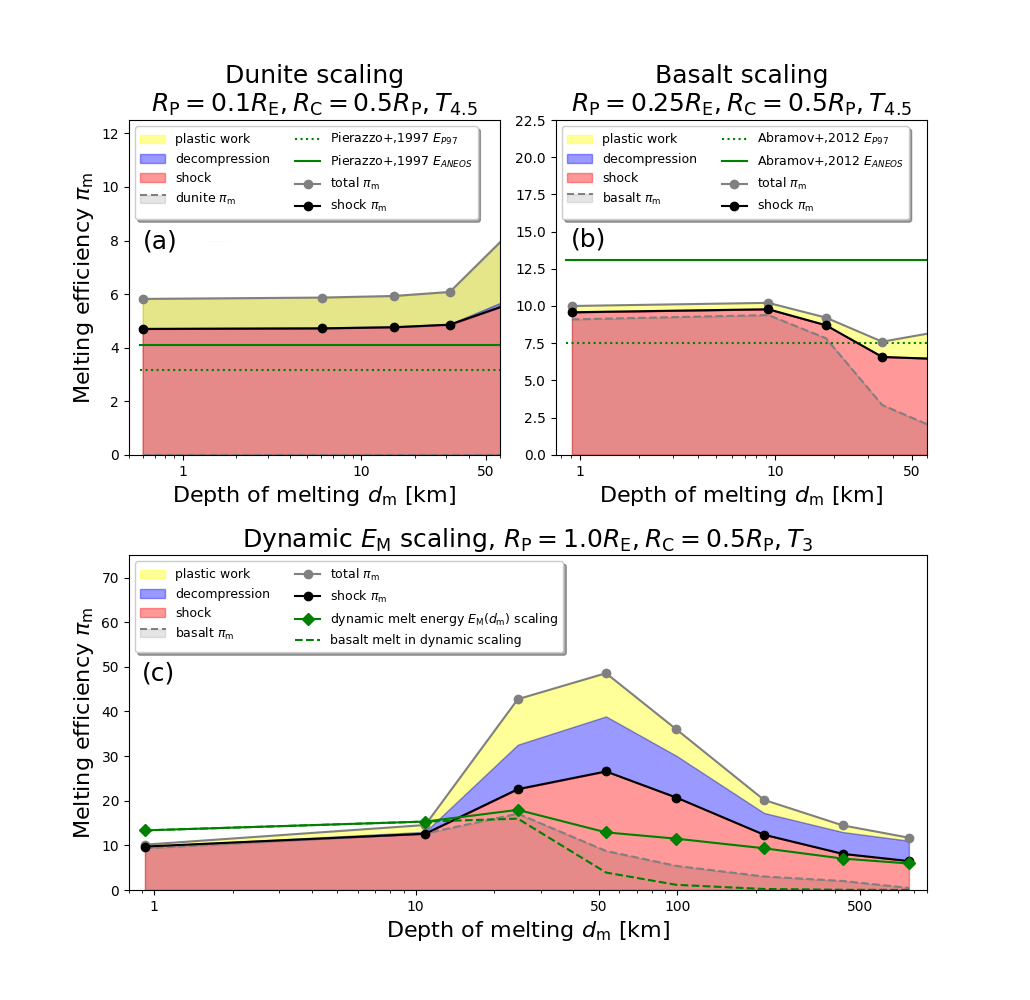}
    \caption{Comparing melting efficiency scaling laws from the literature with our data. In the upper panels, we compare scaling laws under homogeneous target conditions with the shock melting quantities in our results for impacts into a purely dunitic (a) and a mostly basaltic (b) target. The scaling laws are applied using two different values for the specific internal energy of melting $E_\mathrm{M}$: (1)  $E_\mathrm{M}$ is calculated via ANEOS ($E_\mathrm{ANEOS}$, cf. main text). (2) $E_\mathrm{M}$ is taken from the literature and is set to $E_\mathrm{P97} = 9$~MJ/kg for dunite after \citet{Pier:etal97} and also for basalt after \citet{ABRAMOV2012906}. In (c) we compare our data with these scaling laws in a modified way, taking into account the liquidus curve and the initial temperature profile $T_{\mathrm{0}}(z)$, which translates into an increased specific internal energy $E_{\mathrm{0}}(z)$.}
    \label{fig:melteff-scaling}
\end{figure}

Previous studies developed melting efficiency parameterizations, so-called ``scaling laws'', to estimate the melt production for given impactor and target properties. In our setup, melt production in the crust melting regime, which results from relatively small impacts into homogeneous crustal material, can be predicted properly by such scaling laws. In Figure~\ref{fig:melteff-scaling} we plotted the melting efficiency predicted by scaling laws for basalt \citep{ABRAMOV2012906} and dunite \citep{Pier:etal97} and compare it with our model results. Scaling laws are only capable of accounting for shock melting and are usually designed for impacts under homogeneous target conditions (e.g., room temperature and no significant lithostatic pressures). Hence, we compare the scaling laws (green solid line) with the shock melting efficiency data of our models (black solid line) involving relatively small impacts into homogeneous dunitic (a, $R_\mathrm{P}=0.1 R_\mathrm{E}, R_\mathrm{c}=0.5 R_\mathrm{P}, T_\mathrm{4.5}$) and homogeneous basaltic targets (b, $R_\mathrm{P}=0.2 R_\mathrm{E}, R_\mathrm{c}=0.5 R_\mathrm{P}, T_\mathrm{4.5}$, for $d_\mathrm{m}\leq10$~km). To this end, we calculated the specific internal energy of melting, which is defined by the energy of the Rankine--Hugoniot state from which the adiabatic release reaches the liquidus at 1~bar \citep{BjHo87}. Under these conditions we find the specific internal energy of melting $E_\mathrm{M} = 4.7$~MJ/kg for basalt and $E_\mathrm{M} = 7.1$~MJ/kg for dunite in our models using ANEOS and the liquidus functions and apply them in the scaling laws. For dunite, the scaling law fits the data well under approximately constant target conditions ($d_\mathrm{m} \leq 30)$. Compared to the scaling law, the shock melting efficiency in our model is elevated slightly due to additional partial shock melt that was only produced in combination with plastic work \citep[also see][]{Mans:etal22}. The scaling law for basalt, however, differs more strongly from the model data, possibly because it was not derived from parameterized hydrocode simulations and because the chosen value of the specific melting energy $E_\mathrm{M}$ is approximated by adopting the ANEOS-based calculations from \citet{Pier:etal97} for dunite. 

To account for more complex target conditions in scaling laws, \citet{ABRAMOV2012906} and \citet{de2016impact} introduced and demonstrated a modified scaling law in which the specific melting energy is reduced as a function of melting depth, which in turn depends on the liquidus curve and the thermal profile $T_{\mathrm{0}}(z)$; furthermore, they account for latent heat, which unfortunately is not possible in our shock physics code simulations due to restrictions by ANEOS in combination with the solidus and liquidus functions. To compare our data with such a dynamic scaling law, we adopt this approach by calculating the average specific melting energy $\overline{E_\mathrm{M}}(d_\mathrm{m})$ within the melting zone for individual planets and impactor sizes. To this end, we calculate $\overline{E_\mathrm{M}}$ from material within the melting zone which is defined by a spherical volume with a radius of $d_\mathrm{m}$, centered at the depth of melting below the impact point. Note that the melting energy is effectively reduced along the depth profile due to the variations in the liqudus function and by the pre-impact specific internal energy $E_0(z)$; for instance, $E_0$ is significantly elevated within deeper seated, hot material. Normally, the initial energy $E_0(z)$ is small and can be neglected at homogeneous target conditions (see above). However, as this assumption is not valid for the large impacts considered here, we included it in the calculation of the effective specific internal energy $E_\mathrm{M,eff}(z)$, which is defined  as $E_{\mathrm{M,eff}}(z)=E_{\mathrm{M}}(T_{\mathrm{l}}(z))-E_0(z)$, i.e., $E_{\mathrm{M,eff}}(z)$ is the difference between the specific energy from which the adiabatic release reaches the liquidus curve $T_{\mathrm{l}}(z)$ at the pre-impact lithostatic pressure $p_0$ and the initial specific energy. Eventually we calculate the average melting energy $\overline{E_\mathrm{M}}$ of the melting zone by weighting the depth-dependent melting energy $E_{\mathrm{M}}(z)$ with the volume $V_z(z)$ of a certain depth $z$ inside the sphere:
\begin{equation}
\overline{E_\mathrm{M}}(d_\mathrm{m})=\frac{3\pi}{4d_\mathrm{m}^3}\int\limits^{2d_\mathrm{m}}_0 E_{\mathrm{M,eff}}(E_0,T_{\mathrm{l}},z) V_z(z)\,\mathrm{d}z.
\end{equation}
By incorporating the specific melting energy $\overline{E_\mathrm{M}}(d_\mathrm{m})$ into the previously introduced scaling laws for dunite and basalt, we can compare it with the melting efficiency data from our models as shown in Fig.~\ref{fig:melteff-scaling}c, where the dynamic scaling law is plotted alongside melting efficiency data for $R_\mathrm{P} = R_\mathrm{E}$, $R_\mathrm{c} = 0.5 R_\mathrm{P}$ and $T_3$ as an example. For very large and very small impactors, i.e., in the crust (i) and mantle (iii) melting regime, the predicted melting efficiency data are reasonably close (according to the individual scaling laws) to the shock proportion of the melting efficiency from the model data. However, the transitional melting regime (ii) is not represented well by the traditional analytical laws. The scaling law cannot predict peaks in melting efficiency by design and the dynamic scaling law may fail by underestimating the peak in shock melting efficiency, as demonstrated in Figure \ref{fig:melteff-scaling}. Besides the simplified nature of the scaling law, the latter can be explained by additional shock melt in the model, which is produced in combination with the other melting mechanisms, not being taken into account by scaling laws. This extra shock melt pertains to the shock-related part of additional melt that would not have been formed without the effect of the additional melting mechanisms, because the shock pressure alone was not high enough (cf. Fig. \ref{fig:Tz-RK}, shock with and without plastic work). However, we show that scaling laws, even if they account for a depth-dependent melting energy, cannot accurately predict melting efficiency on scales larger than those in the crust melting regime (i), if at all (especially if the crust material is sensitive to plastic work melting), mostly because scaling laws cannot predict the effects of decompression and plastic work melting.

\citet{Marc:etal14} also made an attempt to take the layered structure of the target into account by distinguishing between a crust (granitic in their case) and a dunitic mantle, which are assigned different critical shock melting pressures. Similar to our results, they observed deviations from the strict power-law trends of traditional scaling laws in melt production with increasing impactor size that are caused by the depth variations in target lithology and temperature. In particular, their calculation with the iSALE temperature method in their Extended Data Fig.~4 displays a maximum for intermediate impactor sizes in a range that overlaps partly with that of our models if their melt volume is normalized with the impactor volume, although the peak is much less pronounced in their models than in ours. Its smaller amplitude may be due to an artificial reduction of melting as a consequence of numerical diffusion of the temperature field in their models, which has been addressed in more detail by \citet{Mans:etal22}. They do not, however, provide a scaling law or some other sort of functional expression that quantifies melt production.

\begin{figure}[H]
    \centering
    \includegraphics[width=\textwidth]{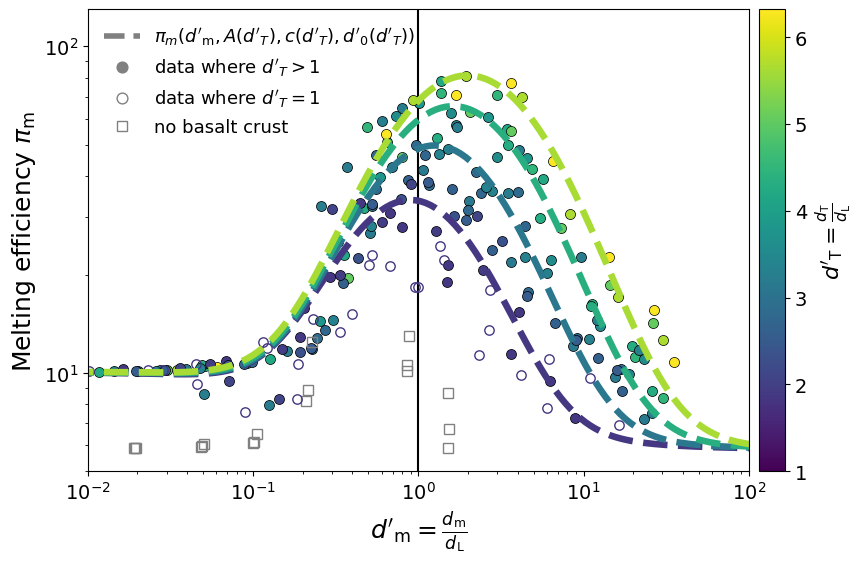}
    \caption{Melting efficiency fitted as a function of $d'_\mathrm{m} = d_\mathrm{m}/d_\mathrm{L}$ shown by dashed lines for different values of $d'_\mathrm{T}= d_\mathrm{T}/d_\mathrm{L}$ (1, 2.5, 4, 5.5), as indicated by the color. The fitted data are shown as filled dots.}
    \label{fig:melteff-univ}
\end{figure}

We aim at constructing such a function that captures the multiple factors that control melt production in a concise, easily applicable form. To this end, we introduce an empirical function by fitting our data as a first step towards improved melting efficiency scaling laws: Fig. \ref{fig:melteff-univ} illustrates the relationship between melting efficiency $\pi_\mathrm{m}$ and the depth of melting, normalized by the lithosphere thickness, denoted as $d'_\mathrm{m} = d_\mathrm{m}/d_\mathrm{L}$. The entire data are shown by different markers. Irrespective of the specific parameters under consideration, a consistent observation emerges: the maxima of melting efficiency predominantly occur at a depth of melting $d_\mathrm{m}$ that is approximately equal to the thickness of the lithosphere $d_\mathrm{L}$ or exceeds it by up to a factor of $\sim3$.
We further find a pronounced enhancement of the amplitude of the melting efficiency maximum corresponding to an increase in the ratio of the supersolidus depth, normalized by the thickness of the lithosphere, $d'_\mathrm{T}=d_\mathrm{T}/d_\mathrm{L}$. This ratio balances the two most relevant length-scales that enhance and hinder the peak amplitudes in melting efficiency, i.e., the thickness of the ``hot'' supersolidus depth relative to the ``cold'' depth of the lithosphere the impactor has to penetrate through. Thus, elevated values of this ratio correspond to relatively hot thermal profiles and a relatively thin upper thermal boundary layer typical for a young target planet, resulting in relatively large melting efficiency maxima at a depth of melting $d_\mathrm{m}$ two to three times deeper than $d_\mathrm{L}$. This behavior can be attributed to the impactor depositing a substantial portion of its energy within a depth range characterized by favorable conditions for material melting regarding all melting mechanisms. As one proceeds to higher target ages and thus thicker lithospheres and cooler targets, the near- or supersolidus depth interval shrinks and approaches the base of the lithosphere, resulting in a reduced melting efficiency. Notably, the absence of a peak in melting efficiency as the depth of melting approaches the supersolidus depth ($d_\mathrm{m} \approx d_\mathrm{T}$) can be explained as follows: The depth of melting $d_\mathrm{m}$ primarily describes the center of the region where the shock pressure peaks and thus primarily shock melting is induced. While shock melting is the most important melt mechanism, the contribution of decompression and plastic work melting significantly add to melt production. For example, a large contribution from decompression melting can be expected if $d_\mathrm{m}$ slightly exceeds $d_\mathrm{L}$. Here, hot material situated further below $d_\mathrm{m}$ does not undergo melting due to the shock but rather due to decompression during the uplift in the course of crater formation.

We parameterize our numerically determined melting efficiencies as a function of $d'_\mathrm{m}$ and $d'_\mathrm{T}$, as shown in Figure \ref{fig:melteff-univ} as thick, dashed lines. Note that only data have been considered where the thermal profile at some part aligns with the solidus ($d'_\mathrm{T} > 1$). While this approach was chosen for the sake of simplicity, data from ``colder'' thermal profiles cannot be addressed. To fit the three melting regimes of the data, a bell-shaped curve to fit the peak in melting efficiency in regime (ii) and a reversed sigmoidal curve to capture the transition between a basaltic crust and a dunitic mantle of regimes (i) and (iii), respectively, are necessary. We tested different functional models and settled on a Gaussian function for the peak and a hyperbolic tangent function for the sigmoidal part, as the different models yielded quite similar fits. The following fit may be used to estimate the melting efficiency on planets with similar structure and thermal profiles, where $\pi_\mathrm{m}^\mathrm{c} = 10.1$ and $\pi_\mathrm{m}^\mathrm{m} = 5.81$ are the ``cold'' melting efficiencies for basalt (crust) and dunite (mantle), respectively (cf. Fig.~\ref{fig:melteff-scaling}):

\begin{gather}
    \pi_\mathrm{m}(d'_\mathrm{m}) = \pi_\mathrm{m}^\mathrm{base}(d'_\mathrm{m},d'_0,\pi_\mathrm{m}^\mathrm{c},\pi_\mathrm{m}^\mathrm{m}) + \pi_\mathrm{m}^\mathrm{peak}(d'_\mathrm{m},d'_0,A,c)
    \label{eq:fit1}\\
    \pi_\mathrm{m}^\mathrm{base}= \frac{\pi_\mathrm{m}^\mathrm{c} +\pi_\mathrm{m}^\mathrm{m}}{2}  -\frac{\lvert \pi_\mathrm{m}^\mathrm{c} - \pi_\mathrm{m}^\mathrm{m} \lvert}{2} \cdot \tanh\left(\lg\frac{d'_\mathrm{m}}{d'_0}\right)
    \label{eq:fit2}\\
    \pi_\mathrm{m}^\mathrm{peak}= A\exp\left[-\left(\frac{\lg \frac{d'_\mathrm{m}}{d'_0}}{c}\right)^2\right],
    \label{eq:fit3}
\end{gather}
where the fitting parameters $d'_0$, $A$, and $c$ can be written as linear functions of $d'_\mathrm{T}$ as long as $d'_\mathrm{T}>1$:

\begin{gather}
    d'_0=0.2432d'_\mathrm{T}+0.6791\\
    A=10.5774d'_\mathrm{T}+15.3064\\
    c=0.023d'_\mathrm{T}+0.5364.
\end{gather}
Further details on the fit are given in the Appendix.
Future attempts to parameterize melt production should incorporate the thermal profile and the solidus and liquidus function, combined with the corresponding equation of state, which also accounts for latent heat.

We emphasize that the pronounced form of the productivity peak is due to the fact that $\pi_\mathrm{m}$ is normalized with the size of the impactor; the absolute melt volume does not feature a maximum and shows only a comparably modest ``hump'' in a log-log plot. However, \citet{Mans:etal21a} already demonstrated that for impactors within the range of the peak, there is an excess of melt that would make craters within this size range particularly prone to overflow with melt and be drowned in it. By contrast, smaller or larger craters would in principle appear more distinctly as craters or basins in the surface morphology of the target. On the other hand, larger impact structures tend to overflow naturally since the produced melt volume scales up significantly more than the (transient) crater volume with increasing impactor size \citep[cf.][]{Mans:etal21a}.

Furthermore, we emphasize that our data and the derived scaling law should not be extrapolated to impactor sizes whose length-scale is comparable to the target planet, e.g., to the Moon-forming event. Firstly, the largest possible melting efficiency $\pi_\mathrm{m,max} = (V_\mathrm{P}+V_\mathrm{imp})/V_\mathrm{imp}$ is limited by the planet's volume. Secondly, the cores of the planet and, potentially, of the impactor would have to be considered. Furthermore, the displacement of the material in such large collisions can differ substantially from the crater formation processes observed in our data, which may lead to significant differences in the resulting decompression melting. In extreme cases, the collisions are disruptive \citep[e.g.,][]{genda2015resolution} and the target planet loses mass as a result of the collision.

\subsection{Global melt production due to impactor flux}
So far, we have assessed melt production for individual impact events across a spectrum of terrestrial planetary bodies and diverse impactor sizes. However, to gain a more comprehensive understanding of the implications for the total melt production in individual planets throughout their evolution, we determine the cumulative melt volume generated by a given impactor flux. In order to put our considerations on an established basis, we assume an impactor flux similar to the well-studied flux the Moon experienced. We calculate the accumulated melt volume for a fixed impactor velocity of 15 km/s as well as for a dynamic impactor velocity that takes the gravity of the individual planet into account. Furthermore, gravitational focusing is considered, which effectively increases the surface area of each planet, depending on its gravity (cf. Sect.\ref{subsect:impvel}). We consider a time period of 4.5 Gyr corresponding to the evolution timescale of the generic planets. In this process, we combine our melt data with a time-dependent impactor flux. To achieve this, we combine the Neukum Production Function (NPF), which describes the cumulative production and abundance $N(D_\mathbf{cr})$ of craters larger than a given final crater diameter $D_\mathbf{cr}$ at a specific period of time on the lunar surface \citep[e.g.,][]{neukum2001cratering,ivanov2001mars}, with scaling laws \citep{holsapple2007crater} to convert $D_\mathrm{cr}$ into an impactor size $L$, resulting in an impactor flux function $N(L)$. This function can then be combined with melting efficiency data $\pi_\mathrm{m}(L)$. To include the largest planets in this study, we had to extrapolate the NPF for $D_\mathrm{cr} > 2300$~km with $\lg N = 5.1398 - 4.6589 \lg D_\mathrm{cr}$. To scale the velocity with the planet's individual escape velocity $v_\mathrm{esc}$, we used data from \citet{Mans:etal21a} with a similar target structure in a velocity range of 10--20 km/s. By incorporating these factors, we aim at gaining insights into the overall melt production and its implications for planetary evolution. A detailed description of how the velocities were calculated and scaled can be found in the Appendix.

\begin{figure}[H]
    \centering
    \includegraphics[width=\textwidth]{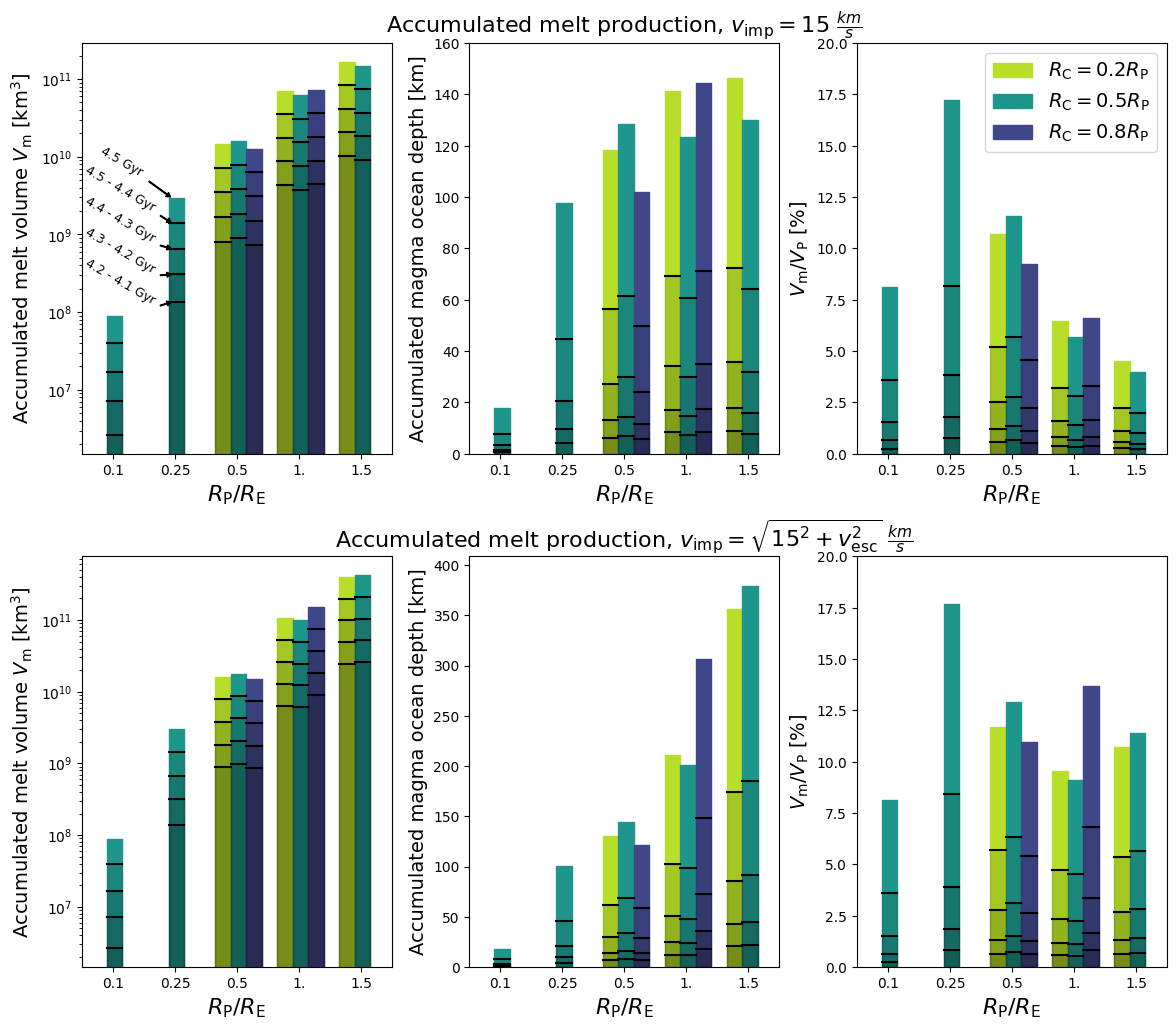}
    \caption{Accumulated melt production over a 4.5 Gyr time interval based on the lunar impactor flux. Here the accumulated melt volume $V_M$, its conversion into a uniform magma ocean depth and the melt volume normalized by the planet volume $V_\mathrm{M}/V_\mathrm{P}$ are plotted. In the bars, different color shading separated by black lines indicates the melt production accumulated over the full 4.5 Gyr and smaller time intervals, as indicated for a Moon-size planet in the upper left panel.}
    \label{fig:melt-pl-size-total}
\end{figure}

Figure \ref{fig:melt-pl-size-total} illustrates the cumulative melt volume and the melt volume relative to the planet's volume as a function of planet size and structure for a time interval corresponding to the lunar impactor flux since 4.5 Ga. The melt of individual impacts calculated here is accumulated according to the impactor flux. In this calculation, the melt volume gives an upper bound, since cooling and crystallization of the melt and impacts into magma oceans, which produced vapor rather than melting deep-seated high-pressure material, are not considered for the sake of simplicity. Assuming a lunar impactor flux, we find that cumulative melting normalized by the planet's volume is most efficient for Moon-size planets, although cooling processes may also be very efficient here. Smaller planets or planetesimals ($R_\mathrm{P} = 0.1 R_\mathrm{E}$) are too cold and do not receive a sufficiently large impactor flux to produce large quantities of melt. On larger planets, large impactors are not efficient enough to melt deep-seated high-pressure material, presumably because of the steeper increase of the lithostatic pressure with depth and the resulting steeper slope of the melting curves. When projecting these findings to the Solar System, one may argue that the Moon has produced more melt relative to its size than the Earth. However, melt production also increases with impact velocity. In general, the gravity of the central star dominates $v_\mathrm{imp}$ for close-in stars, especially small ones (cf. Sect.~\ref{subsect:impvel}), but for targets farther than a few au from the star, where the star's diminished influence causes an overall reduction of the effective $v_\mathrm{imp}$, the relative contribution of the targets' own gravity increases and may shift relative melt productivity in favor of larger planets.\par
We note that the most substantial melt volumes are accumulated within the first few hundreds of millions of years. For instance, approximately half of the total melt is produced during the first 100 million years of bombardment. However, we approximate the planets' thermal evolution using 1, 3, and 4.5 Gyr-old thermal profiles, because it is difficult to calculate profiles directly after planet formation. Additionally, antipodal melting is not considered in our model. However, it occurs at very large impact events where melt is produced on the opposite side of the planet when sufficiently strong shock waves converge at the impact site's antipode. Thus our melt volume estimates can be interpreted as a lower limit, since thermal profiles younger than 1 Gyr would most likely result in the production of more melt. Furthermore, very large impacts change the gravity field of the target planet, modifying its lithostatic pressure and the material displacement throughout crater formation, which in turn affects decompression and plastic work melting. These effects are not captured in our model either, because we use a central-gravity model in which the gravity is set once at the beginning and remains unchanged throughout the run.

\subsection{Impact angle}
\label{subsect:impang}
Due to the expansive parameter space covered by this study with more than 200 impact simulations, we have chosen to limit our analysis to 2D simulations, restricting impact angles to 90\textdegree. The scope of our study precludes more computationally intensive 3D simulations, which do allow for different impact angles that may significantly affect melt production. \citet{pierazzo2000melt} demonstrated that shock-induced melt production is reduced by 20\% and 50\% when altering the angle from 90\textdegree\ to 45\textdegree\ and 30\textdegree, respectively. Notably, an impact angle as shallow as 15\textdegree\ led to a substantial 90\% reduction. Conversely, \citet{wakita2019enhancement} revealed a contrary outcome when accounting for plastic work heating; diminishing the impact angle from 90\textdegree\ to 45\textdegree\ can actually slightly amplify impact-induced heating at an impact velocity of 5 km/s. In a follow-up study, in which a broader range of impact angles and velocities was examined, those authors conclude that total heating, including the effect of shock and plastic work, is not significantly affected by the impact angle unless it drops below 45\textdegree\ \citep{wakita2022effect}. It is worth noting that their investigations both employed dunitic material and an impact velocity of up to 10 km/s, emphasizing the prominence of plastic work melting in this range of rather low impact velocities \citep{Mans:etal22}. For higher velocities, the dominance of shock melting may challenge the conclusions of these studies since plastic work melting becomes negligible \citep{Mans:etal22}, while the influence of impact angle reduction leads to the reduction of shock, and thus potentially total, melting, counteracting this trend.\par
These findings collectively suggest a dual effect of decreasing impact angle: a reduction in shock melting and an enhancement of plastic work melting, at least for impact angles from 90\textdegree\ to 45\textdegree\ and impactors slower than about $10$ km/s \citep[cf.][]{Mans:etal22}. Furthermore, the interplay between impactor angle, shock, plastic work and decompression melting necessitates deeper investigation, especially because these melting mechanisms are affected by the thermal profile and impact velocity \citep{Mans:etal21a,Mans:etal22}. For impacts with shallower angles less than 45\textdegree\, we anticipate a shift of the peak in decompression melting efficiency to larger impactor diameters because of the reduced depth of melting. The intricate relationship among decompression, shock, and plastic work melting in combination with given thermal profiles and impact velocities warrants comprehensive examination to elucidate the extent of their interdependence. However, a substantial reduction of melt production for impact angles below 45\textdegree\ seems to be very likely. 

\subsection{Impact velocity}
\label{subsect:impvel}
We have limited ourselves to a single impactor velocity, 15 km/s, a choice inspired by an estimate by \citet{Chyba91} of the expected average velocity for asteroidal impactors striking the Earth. Given the key importance of this parameter for the energy budget of the impact and hence for its productivity with respect to melt, the implications of this choice deserve some more detailed consideration. How representative is our, or indeed any, choice for impacts on terrestrial planets? To address this question, we begin by pointing out that the mean impact velocity in the stationary reference frame of the target planet is the geometric mean 1.) of the velocity $v_\mathrm{ip}$ of the impactor due to the gravitational potential of the central star of mass $M_\mathrm{s}$ at the distance of the planet and 2.) of its velocity due to the gravitational attraction of the target at its surface, which is in fact its escape velocity $v_\mathrm{esc}=\sqrt{2G_0M_\mathrm{P}/R_\mathrm{P}}=\sqrt{2gR_\mathrm{P}}$ (neglecting rotation), i.e.,

\begin{equation}
v_\mathrm{imp}=\sqrt{v_\mathrm{ip}^2+v_\mathrm{esc}^2}
\label{eq:vimp}
\end{equation}
\citep[e.g.,][]{HuWi00}. The more the apastron distance $d_\mathrm{ap,i}$ of the impactor exceeds the radius of the planetary target's orbit $a_\mathrm{P}$, the more eccentric must its elliptical orbit be for a collision to be possible in the first place. For the following estimates, we assume circular orbits around a Sun-like star with $a_\mathrm{P}$ of 0.1, 1, 10, and 30 au for the target planets and use the formalism by \citet{Opik51} or \citet{HuWi00} as outlined in \citet[App. A]{RuBr18b} to estimate $v_\mathrm{imp}$ for impactors with $d_\mathrm{ap,i}$ between $a_\mathrm{P}$ and 40 au. As $M_\mathrm{s}\gg M_\mathrm{P}$, $v_\mathrm{imp}$ is usually dominated by the gravity of the central star, and in general the more strongly so the closer the target planet is to the star, the smaller the target is, and the more eccentric and/or the more steeply inclined the orbit of the impactor is. As a consequence, at small inclinations of the impactor our nominal impactor velocity of 15 km/s occurs frequently in low-eccentricity impactors at small close-in planets, but beyond a few au it is not reached anymore, because the gravitational pull of the central star is already too weak, and the mass of the target itself is not sufficient. In those remote realms, it is therefore rather the larger planets that can accelerate impactors to 15 km/s on their own, whereas large close-in planets almost always feel much higher $v_\mathrm{imp}$. Especially in large planets at greater distances from the central star, where the gravitational pull of the planet is important or even dominant, the size of the core can have a substantial influence on $v_\mathrm{imp}$ due to its large effect on the mass of a planet of given size.\par
In conclusion, we consider our choice of $v_\mathrm{imp}$ a reasonable compromise if one has to limit oneself to a single value, but we note that for close-in and/or large planets with a given internal structure it will be rather too low, and melt production will be underestimated, whereas it is rather too high for small far-out planets, where melt production will thus be more muted in comparison with our models. Apart from the vertical impacts of our models, the surface-normal component of the impact velocity may of course be close to 15 km/s even in high-$v_\mathrm{imp}$ impacts, but in such cases the melt distribution will not be axially symmetric. Moreover, melt production can be extrapolated within a certain range to slower or faster impact velocities as demonstrated and discussed previously and in the Appendix~\ref{sect:v-scaling}.\par
The impact velocity also controls the velocity of ejected material and thus whether and how much of it can be lost permanently. As explained in more detail in Appendix~\ref{sect:ejecta}, only a small fraction of the ejected mass in the cases considered here and in many other situations is expected to be accelerated beyond the escape velocity of the target. Therefore, it is safe for us to neglect the permanent loss of ejecta and in particular of ejected melt in our considerations.

\subsection{Some implications for the Solar System and for exoplanets}
The models investigated here are generic planet models, but the choice of their internal structures was obviously inspired by the planetary bodies of the Solar System. Therefore, we will briefly discuss some possible implications of the models for impact processes on the planets from which they were derived. Such implications, however, cannot be very specific because of the highly idealized model setup. Important idealizations include the restriction to vertical impacts, which enforces cylindrical symmetry on the entire model, and the limitation to a single impact, which implies that consecutive impacts do not influence their successor events, or in other words, that the geodynamical evolution of a planet is not affected by the memory of earlier impacts in the geological record. Fully dynamical models of long-term mantle evolution demonstrated already that the prolonged succession of many impacts has a noticeable cumulative effect on the evolutionary path of a planet via the perturbations of its heat budget and compositional structure \citep[e.g.,][]{RuBr19a}.

For instance, we observed faster secular cooling and, as a consequence, a reduction of melting efficiency by as much as 60\% in small planets. Applied to Mercury and the Moon, this suggests that melting in late large impacts on them should be less productive than on similar-sized impacts on larger planets. A comparison between craters from similar impacts on the Moon and on cratons on Earth would be instructive due to the similar mean impact velocities and the relatively small effect of relative core size for bodies with small to medium-size cores; cratons probably come closest to our modeling assumption of a stagnant lid, whereas the evolution and properties of younger types of lithosphere on Earth would be affected by the combination of their young geologic age and the long-term changes that have occurred globally in a much older mantle. Mercury is more difficult to compare with other planets, because a decline could in part also be due to the effect of the large core but would on the other hand be masked by the much higher impact velocities there.

Venus does not seem to have a strictly stagnant lid, but it is not in a fully mobile plate-tectonics regime like Earth either. Thus, some results from our generic planets may be applicable to it. Comparisons should probably be limited to the young-planet models, because Venus's surface has been rejuvenated less than 1 Gyr ago and because much of its lithosphere seems to be quite thin at 20 km or less \citep{FSAnSm06}. An impact on modern Venus may therefore correspond more to one on our 1 Gyr-old Earth-like model and should thus display a high melting efficiency, higher than would otherwise be expected from a planet of its age. Moreover, its high surface temperature~-- much above the 288 K of our models~-- put even the shallowest parts of the crust much closer to the solidus than is the case in our models, which will reinforce melting efficiency further. Much of the melt production would likely be due to shock and decompression, as thermal weakening of the hot lithosphere should reduce the plastic work contribution. In principle, these conditions could make even small impacts unusually productive, but the dense atmosphere filters out small impactors.

We do not discuss the results for the smallest planet size included in our models ($R_{\mathrm{P}} = 0.1 R_{\mathrm{E}}$), because such small planets exhibit a different dynamical behavior than larger planets, as they are too cold and small to initiate mantle convection, resulting only in a conductive thermal profile. Consequently, the mantle is comparably cold, and the capacity to build a crust through volcanism at least in the early stage of evolution mostly depends on the concentration of radionuclides whose decay provides internal heating. Our data suggest that this difference in internal dynamics appears at a threshold planet size of $0.1 R_\mathrm{E}\lesssim R_{\mathrm{P}} \lesssim 0.25 R_\mathrm{E}$, where melt production is substantially decreased. Applied to the large, differentiated asteroids in the Solar System, it implies that melt formation by impacts is much less important on them than it is on planets. The generally lower impact velocities on such small bodies also contribute to the reduction of melting efficiency.

An important outcome of these models is how strongly melt generation in an impact is controlled by the combined effects of the impact parameters and the internal structure of the target planet. The latter is reasonably well known for the terrestrial bodies in our Solar System, but for the application of this method to terrestrial-type exoplanets, it must be kept in mind that the results are subject to the additional uncertainties introduced by the lack of information about their internal structure.

\section{Summary and Conclusions}\label{sec:conclusions}

We analyzed impact-induced melt production on various generic planets with different thermal and evolutionary states, core and planet sizes, and crustal thicknesses at an impact velocity of 15 km/s. The results show that as the planet's interior cools, the amount of melt produced by impacts decreases substantially, with a reduction in melting efficiency by about 40\% for large to 60\% for small planets throughout their evolution from 1 Gyr to 4.5 Gyr after formation. Peaks in melting efficiency occur at impactor sizes where the depth of melting $d_{\mathrm{m}}(L)$ is similar to or up to about three times greater than the thickness of the lithosphere $d_{\mathrm{L}}$. We found that the contribution of decompression and plastic work melting to total melt production becomes significant as the impactor size increases, although shock melting generally remains the dominant mechanism. Plastic work melting can contribute up to 20--30\% of the total melt. Decompression melting was always enhanced on hotter planets and tends to increase with increasing impactor size up to almost 50\% at an impactor size of $L = 1000$ km.\par
Furthermore, we find the following for sufficiently large $d_\mathrm{m}$ where the impact penetrated through the planets upper, relatively cold surface layer: for larger planets, smaller impactors are more efficient at producing melt, while for smaller planets, larger impactors are more efficient. This is because smaller planets have a thicker outer thermal boundary layer and a smaller increase in lithostatic pressure with depth, which results in a deep-seated layer where the geotherm is close to the solidus (``supersolidus''), making it easier to melt with larger and deeper penetrating impactors. By contrast, larger planets have a steeper lithostatic pressure gradient but a thinner thermal boundary layer, which leads to a large temperature difference between the geotherm and solidus with increasing depth, making smaller impactors more effective at producing melt. Thus, maxima in melting efficiency appear at different impactor sizes or melting depths at differently sized planets. When comparing the amplitude of those maxima we find that usually melting is most efficient on Earth-size planets. However, the melting efficiency is not strongly affected by the core size ratio except for very large cores and older planets, where melt production is strongly reduced compared to smaller core size ratios due to more efficient cooling of the mantle during the evolution of the planet.

Our results shed light on the general mechanisms behind impact-induced melt production and their implications for planet formation and evolution. Furthermore, they offer a useful parameterization for predicting the amount of melt that can be produced on planets with similar sizes, internal structures, and internal temperatures. By combining our findings with an impactor influx function \citep[e.g.,][]{neukum2001cratering}, we can estimate the total amount of impact-induced melt for a given evolutionary stage. We found that smaller planets, as long as they have a hot thermal depth profile, are more likely to develop large melt ponds or even small magma oceans than larger planets with the same impactor flux, even when considering gravitational focusing and increased impact velocities for larger planets and cores due to their increased gravity. Projecting our findings onto the Solar System, this would mean for instance that impact-induced melting on the Moon and possibly also on Mars used to be more pronounced than on Earth.

This work also highlights the importance of considering the contribution of decompression and plastic work melting relative to shock melting. Existing scaling laws for predicting impact-induced melting often only account for shock melting, but we demonstrate that this approach may not accurately estimate melt for impactor sizes where the depth of melting exceeds the thickness of the lithosphere. Even for smaller impacts, especially with an impactor velocity below 12.5 km/s, plastic work must not be neglected \citep[e.g.,][]{KKuGe18,Mans:etal22}.

\section{Open Research}

The iSALE shock physics code is not entirely open-source but can be distributed on a case-by-case basis to academic users in the impact community for non-commercial use. Readers interested in using the code find the application requirements at the iSALE website (\url{http://www.isale-code.de/redmine/projects/isale/wiki/Terms_of_use}). Furthermore, any iSALE version includes the ANEOS package. The modification in iSALE and the input files to generate the results supporting the figures are described in \citet{Mans:etal22} and \citet{9KKNFE_2024}, respectively.

\section*{CRediT statement}
Conceptualization: T.R.;
data curation: L.M., A.-C.P.;
formal analysis: L.M., T.R.;
funding acquisition: K.W., T.R., N.T., N.A.;
investigation: L.M., T.R., A.-C.P.;
methodology: L.M., T.R.;
resources: K.W., A.-C.P.;
software: L.M., P.B., N.T., K.W.;
supervision: T.R., K.W.;
validation: L.M., A.-C.P.;
visualization: L.M., T.R.; 
Writing – original draft: L.M., T.R., A.-C.P.;
Writing – review \& editing: all authors

\section*{Acknowledgments}
We thank two anonymous reviewers and editors for helpful comments that improved this manuscript substantially.
L.M. and K.W. were funded by the German Research Foundation (DFG, SFB TRR-170-1 TP C2 and C4).
T.R. was partly funded by grant RU 1839/2 from the DFG.
A.-C.P. gratefully acknowledges the financial support and endorsement from the DLR Management Board Young Research Group Leader Program and the Executive Board Member for Space Research and Technology.
P.B. and N.T. acknowledge support of the DFG through the priority program SPP 1992 ``Exploring the Diversity of Extrasolar Planets'' (TO 704/3-1) and through the research unit FOR 2440 ``Matter under planetary interior conditions'' (PA 3689/1-1).
N.A. was supported by an Alexander von Humboldt research award. This is TRR~170 contribution 225. We gratefully acknowledge the developers of iSALE-2D, including Gareth Collins, Dirk Elbeshausen, Tom Davison, Boris Ivanov, and Jay Melosh.
Open Access funding enabled and organized by Projekt DEAL.

\appendix

\section{Impact velocity scaling}
\label{sect:v-scaling}
The impact simulations in this study are conducted at a single impact velocity of 15 km/s. However, to account for the gravitational attraction of the planet as given by Eq.~\ref{eq:vimp}, we rescaled the impactor velocity assuming a velocity due to the potential of the central star of $v_\mathrm{ip} = 15$ km/s. For this purpose, we used data from \citet{Mans:etal21a} where melting efficiency has been determined for a Mars-like thermal profile with a basaltic crust and a dunitic mantle for impact velocities of 10, 15, and 20 km/s, considering decompression and shock melting. 

\begin{figure}[H]
    \centering
    \includegraphics[width=\textwidth]{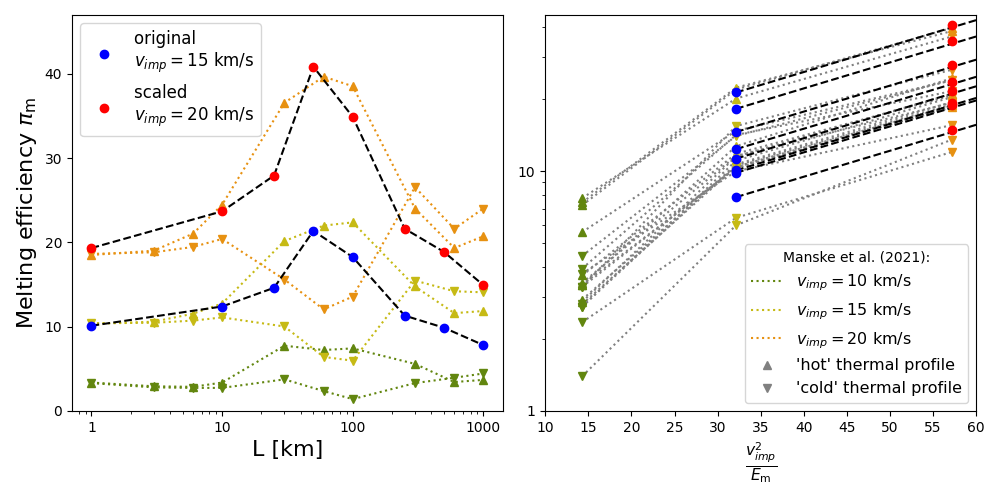}
    \caption{Melting efficiency as a function of the melt number $v_\mathrm{imp}^2/E_\mathrm{M}$ and impactor diameter $L$ scaled by using data from \citet{Mans:etal21a} as indicated by dotted lines and triangles. The colors of the dotted lines and triangles indicate the impact velocity corresponding to the data (orange for $v_\mathrm{imp}=20$ km/s, olive for $v_\mathrm{imp}=15$ km/s, and green for $v_\mathrm{imp}=10$ km/s) and triangles facing upwards and downwards indicate the corresponding data points for hot and cold thermal profiles, respectively. As an example, data for an 4.5 Gyr-old Earth-size planet with a large core has been scaled from 15 km/s (blue dots) to an impact velocity of 20 km/s (red dots). Both legends are valid for both plots.}
    \label{fig:melt-v-scaling}
\end{figure}
\citet{Pier:etal97} found that the impact-induced melt volume can be scaled with respect to velocity by a power law. We follow their approach and scale the melting efficiency data from \citet{Mans:etal21a} with the melt number ($v_\mathrm{imp}^2/E_\mathrm{M}$) by calculating the average slope in melting efficiency between the corresponding data with different impact velocities. Here $E_\mathrm{M} = 7.1$~MJ/kg is the specific melting energy required for melting of dunite under atmospheric pressure and has been determined by comparing ANEOS with the liquidus $T_\mathrm{l}$. Since the scaling approach only holds true for $v_\mathrm{imp}^2/E_\mathrm{M} > 30$ \citep{Pier:etal97}, we scaled only data with $v_\mathrm{imp} \geq 15$~km/s, which anyways fits the velocity range of interest. Figure \ref{fig:melt-v-scaling} displays the data from \citet{Mans:etal21a} (triangles) used to derive the slope of the power law to scale the data of this study. In the figure the melting efficiency has been scaled for an Earth-sized planet with a large core from 15 km/s (blue dots) to 20 km/s (red dots). Here, the scaling produces reasonable results when compared to the original data. In the scope of this study we scale the impact velocity $v_\mathrm{imp}$ up to 23.85 km/s as a consequence of the the gravitational attraction of the planet (cf. Table \ref{tab:Vscaling}). The scaling has been applied in Figure \ref{fig:melt-pl-size-total} to account for the additional acceleration due to the gravitational attraction of the planets. The scaling relation after \citet{Pier:etal97} is given by 

\begin{equation}
\lg(\pi_\mathrm{m}) = a + b \cdot \lg \left( \frac{v_\mathrm{imp}^2}{E_\mathrm{M}} \right).
\label{eq:vimp_scale}
\end{equation}

 We determined the slope of the power law as $b = 0.986$ using data from \citet{Mans:etal21a}. The constant $a$ has to be adjusted according to the input data. As the data from \citet{Mans:etal21a} have a similar setup, account for decompression melting, and use a similar range of impact velocities, we expect that the velocity-scaled melting efficiency gives reasonable results. Whether this scaling can be applied to other data may be strongly dependent on the impact and target conditions.

\begin{table}
\caption{The surface gravity $g$, the escape velocity $v_\mathrm{esc}$, and the impact velocity $v_\mathrm{imp}$, calculated in Eq.~\ref{eq:vimp}, are stated for $v_\mathrm{ip}$ = 15 km/s and the different planets at 1 Gyr.}\label{tab:Vscaling}
\begin{tabular}{lcccc}
    \hline
    $R_\mathrm{P}$ &  $R_\mathrm{c}/ R_\mathrm{P}$ & $g$ (m/$\mathrm{s^2})$ & $v_\mathrm{esc}$ (km/s) & $v_\mathrm{imp}$ (km/s)\\\hline
    640 km        & 0.5 & 0.69 & 0.94  &  15.03\\
    1600 km & 0.5 & 1.71 & 2.34  & 15.18\\
    3200 km  & 0.2& 2.98 & 4.37 & 15.62\\
    3200 km& 0.5  & 3.59 & 4.79  & 15.75\\
    3200 km& 0.8 & 5.67 & 6.03 & 16.17\\
    6400 km& 0.2 & 7.25 & 9.63   & 17.83\\
    6400 km& 0.5 & 9.17 & 10.83   & 18.50\\
    6400 km& 0.8 & 15.88 & 14.26   &  20.70\\
    9600 km& 0.2 & 13.61 & 16.16   & 22.05\\
    9600 km& 0.5 & 17.90 & 18.54   &  23.85\\
    \hline
\end{tabular}
\end{table}

\section{Solidus/liquidus parameterizations}\label{sect:Tslpar}
\subsection{Chondritic peridotite}
\begin{figure}
    \centering
    \includegraphics[width=\textwidth]{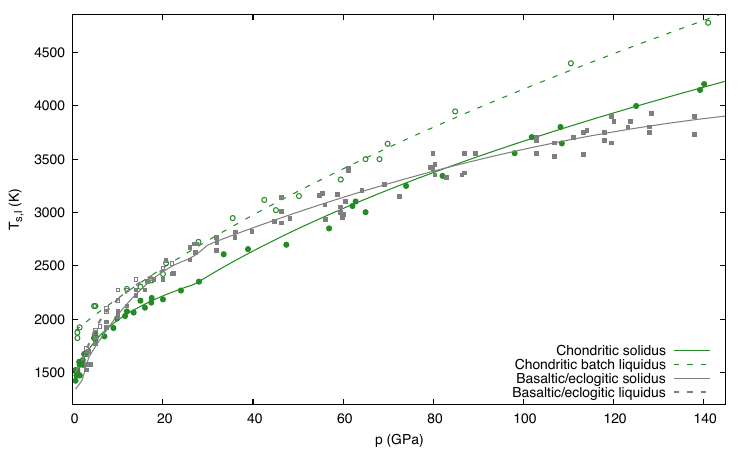}
    \caption{Solidus and liquidus fits for the chondritic and basaltic/eclogitic compositions, and the underlying experimental data. The eclogitic liquidus is only plotted up to 22 GPa.}
    \label{fig:solliq}
\end{figure}

The solidus and liquidus parameterizations for chondritic material, which stands in for the ``dunite'' material in the melting context of the iSALE/ANEOS impact modeling framework, were constructed by fitting data from melting experiments on various natural and synthetic ``chondritic'' compositions, excluding carbonaceous chondrite compositions; for the high-pressure ($p>27.5$ GPa) bridgmanite--ferropericlase solidus, we rewrote the fit from \citet{Andr:etal18} in a form compliant with Eq.~\ref{eq:TslSG}. The solidus fit implemented here follows \citet{Andr:etal18} in treating the data for the (terrestrial) upper and lower mantle pressure range separately. For the latter, only the data from \citet[synthetic CMASF]{Andr:etal11} are available, as a single additional point from \citet{OhSa87} used a simplified (CMASF) composition and is probably not reliable. For lower pressures, we supplemented the extensive dataset from \citet[synthetic CMASFNCrT]{Andr:etal18} with some points from \citet[Indarch EH4]{Bert:etal09}, \citet[synthetic CMASF]{Ohtani87}, and \citet[Yamato Y-74191 L3 chondrite]{Takahashi83} that seem to fit in well. Additional points from \citet[natural and synthetic Homestead]{AgDr04} and \citet[Hvittis enstatitic chondrite EL6]{CCart:etal14} were too far above the general trend to be considered. The liquidus fit uses the datasets by \citet{AgDr04,Andr:etal11,Bert:etal09,Ohtani87} and \citet{Takahashi83}, but the data do not warrant making a distinction between the upper and lower mantle pressure range. The liquidus data are from batch melting experiments and thus represent a system in which the melt does not segregate from the solid phase, as would often be the case in regular mantle melting, but it should be appropriate for the fast melting processes expected in impacts. The solidus and liquidus have data coverage up to about 141 GPa (Fig.~\ref{fig:solliq}), so extrapolation into the stability field of postperovskite is unconstrained and will miss possible deviations arising from the phase change.\par
For technical reasons, we approximated the Simon--Glatzel form of the solidus and liquidus Eq.~\ref{eq:TslSG} with the cubic polynomials

\begin{align}
T_\mathrm{s}&= 1355.08 + 148.17p - 14.946p^2 + 0.6587p^3\\
T_\mathrm{l}&= 1875.44 + 34.376p - 0.26708p^2 + 0.003163p^3
\end{align}
in the thermal evolution models. They are sufficiently accurate in the pressure range of interest for that purpose but should not be used for extrapolation outside the pressure range from 0.25 GPa to 10 GPa (solidus) and beyond 27.5 GPa (liquidus), respectively, to stay within 10 K of Eq.~\ref{eq:TslSG}.

\subsection{Basalt}
For impact melting, the solidus and liquidus of crustal material, here assumed to be basaltic, are also relevant. Basalt consists mostly of clinopyroxene and plagioclase. When the latter transforms into garnet, the basalt becomes eclogite. Data for the solidus of basalt are scarcer than for peridotite or eclogite, and so we interpolate between a few points measured from the (terrestrial) MORB basalt/eclogite phase diagram provided by \citet{LiOh07} for the stability field of basalt, which we extend to 3 GPa in order to achieve continuity with the eclogite parameterization, for which experimental data were fitted. When the eclogite transforms into a bridgmanite-bearing high-$p$ assemblage, the function shifts linearly to the high-$p$ melting curve over a $p$ interval from 26 to 30 GPa. The bound of 26 GPa separates the subsets of experimental data used for fitting the eclogite and the high-$p$ solidus, whereas the 30 GPa bound is somewhat arbitrary. The experimental data are from \citet{Andr:etal14a,Hiro:etal99,HiFe02,Prad:etal15}, and \citet{Yasu:etal94}; from the Andrault et al. data, only those above $\sim56$ GPa are used, because the lower-$p$ data lie significantly beneath those of the other studies.\par
For basalt/eclogite compositions, we found polynomial parameterizations to work better in most cases, and so we used for the solidus the function

\begin{equation}
T_\mathrm{s}(p)=
\begin{cases}
1.83261p^4+13.0994p^2+1340&p<3\,\text{GPa}\\
-1.273p^2+78.676p+1380.92&3\,\text{GPa}\leq p<26\,\text{GPa}\\
-0.0503p^2+19.277p+2165.09&p\geq 30\,\text{GPa,}
\end{cases}
\end{equation}
based on the phase diagram from \citet{LiOh07} and the experimental melting data from the other studies, which cover a pressure range from 3 to 138 GPa. Because of the phase transition to postperovskite at only slightly higher pressure and the polynomial form of the high-$p$ parameterization, substantial extrapolation to higher $p$ is discouraged.\par
The phase diagram from \citet{LiOh07} and some experimental data from \citet{HiFe02} and \citet{Yasu:etal94} are also used to construct a batch melting liquidus for basalt/eclogite in a way similar to the solidus:

\begin{equation}
T_\mathrm{l}(p)=
\begin{cases}
21.5909p+1488&p<0.88\,\mathrm{GPa}\\
-2.24727p^2+97.6753p+1422.79&0.88\,\mathrm{GPa}\leq p<3.12\,\mathrm{GPa}\\
1581.57(p-1.468)^{0.151}&3.12\,\mathrm{GPa}\leq p<22\,\mathrm{GPa}.
\end{cases}
\end{equation}
The interval between 0.88 and 3.12 GPa is derived from \citet{LiOh07} and serves as a link between the lowest-$p$ segment and the range constrained by experimental data. Note, however, that around 22 GPa this liquidus gets very close to the solidus and that it is not at all constrained at higher $p$, where eclogite transforms into a perovskitic assemblage (Fig.~\ref{fig:solliq}); its use at $p\gtrsim 20$ GPa should probably be avoided.

\section{Escape of melt and other ejecta}\label{sect:ejecta}
We did not track whether ejected melt escapes from the target planet, which may be important for high-velocity impacts and low-mass planets. We use the expressions for the velocity of ejected material at the rim of the opening crater from \citet{JERich:etal07} to estimate under which conditions melt intermingled with ejecta is accelerated to velocities above $v_\mathrm{esc}$ and would thus leave the planet for good. Specifically in the gravity regime, \citet[eqs.~41, 29, 30]{JERich:etal07} give the effective velocity as

\begin{subequations}
\begin{align}
v_\mathrm{ef}(r)&=\sqrt{v_\mathrm{e}^2-C_{\mathrm{vp}g}^2gr}=\sqrt{C_{\mathrm{vp}g}^2gR_g\left(\frac{R_g}{r}\right)^\frac{2}{\mu}-C_{\mathrm{vp}g}^2gr}\nonumber\\
&=\frac{\mu}{C_{Tg}(\mu+1)}\sqrt{2g\left[R_g\left(\frac{R_g}{r}\right)^\frac{2}{\mu}-r\right]}\label{eq:vef}
\end{align}
with the radius of the transient crater in the gravity regime given by

\begin{align}
R_g&=\left[\frac{3K_1}{\pi}\frac{m_\mathrm{imp}}{\varrho}\left(\frac{gL}{2v_\mathrm{imp}^2}\right)^{-\frac{3\mu}{2+\mu}}\left(\frac{\varrho}{\varrho_\mathrm{imp}}\right)^\frac{\mu}{2+\mu}\right]^\frac{1}{3}\nonumber\\
&=L\left(\frac{2v_\mathrm{imp}^2}{gL}\right)^\frac{\mu}{2+\mu}\left[\frac{K_1}{2}\left(\frac{\varrho}{\varrho_\mathrm{imp}}\right)^{\frac{\mu}{2+\mu}-1}\right]^\frac{1}{3}
\end{align}
\cite[eqs.~10, 11]{JERich:etal07}; these authors give the constants as $C_{Tg}=0.85\pm0.1$ and $K_1=0.2$ for hard rock. The ejection angle for a vertical impact, measured from the horizontal, is represented by a simple linear function,

\begin{equation}
\psi=\psi_0-\psi_\mathrm{d}\frac{r}{R_g},
\end{equation}
where the constants $\psi_0=60^\circ$ and $\psi_\mathrm{d}=30^\circ$ are determined from numerical models \cite[eq.44, tab.4]{JERich:etal07}.
\end{subequations}
If melt escapes from the planet, the vertical component of its velocity must have at least the escape velocity of the target, i.e.,

\begin{gather}
v_\mathrm{ef}(r)\sin\left(\psi(r)\right)\geq v_\mathrm{esc}\nonumber\\
\Leftrightarrow \frac{\mu}{C_{Tg}(\mu+1)}\sin\left(\psi_0-\psi_\mathrm{d}\frac{r}{R_g}\right)\sqrt{2g\left[R_g\left(\frac{R_g}{r}\right)^\frac{2}{\mu}-r\right]}\geq \sqrt{2gR_\mathrm{P}}\nonumber\\
\Leftrightarrow \sin^2\left(\psi_0-\psi_\mathrm{d}\frac{r}{R_g}\right)\left(\frac{R_g}{r}\right)^\frac{2}{\mu}-\frac{r}{R_g}\geq \frac{R_\mathrm{P}}{R_g}C_{Tg}^2 \left(\frac{\mu+1}{\mu}\right)^2.\label{eq:vef-r}
\end{gather}
The condition would be fulfilled for distances from the impact point smaller than the solution $r_\mathrm{esc}$ of this equation for $r$. On the other hand, it must be kept in mind that for an impactor of finite size, material inside a radius $r_\mathrm{ej,min}=n_1L/2$ is expected to be driven down into the target and not to be ejected; the empirical multiplier $n_1$ is estimated as 1.2 by \citet[Sect.3.1 and 6]{HoHo11} but is relatively poorly constrained. Hence, the region from which ejecta can be accelerated beyond $v_\mathrm{esc}$ is not circular but a ring with inner radius $r_\mathrm{ej,min}$ and outer radius $r_\mathrm{esc}$ centered on the point of impact (in vertical impacts). Although the form of Eq.~\ref{eq:vef-r} guarantees that at some $r$ ejecta would escape from the gravitational ties to the planet, the finite size of the impactor will prevent this from happening unless $r_\mathrm{esc}\geq r_\mathrm{ej,min}$. As a consequence, there are many configurations in which there will be no escape of melt or other ejecta.\par
\begin{figure}
    \centering
    \includegraphics[width=\textwidth]{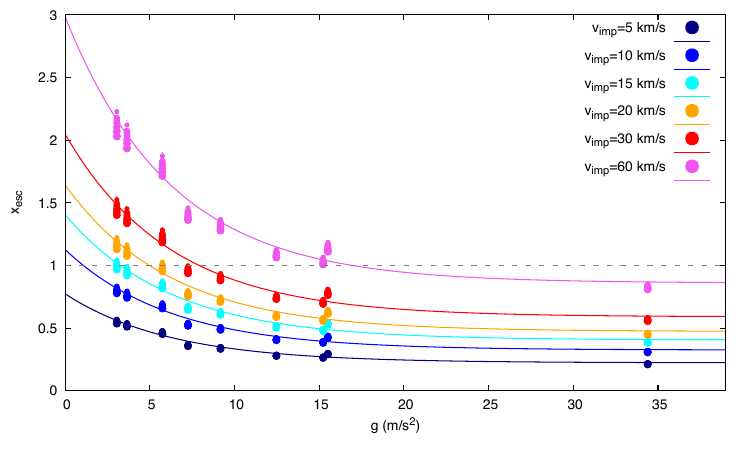}
    \caption{Plots of $x_\mathrm{esc}(g)$ (Eq.~\ref{eq:xesc-fit}) and results for the target planets and impactors of this paper, for different $v_\mathrm{imp}$. $x_\mathrm{esc}$ is the maximum distance from the point of impact at which the vertical component of the velocity of the ejecta surpasses the target's escape velocity, normalized with the minimum distance from which material can be ejected. Escape is thus possible in the cases whose symbols lie above the dashed grey line. Symbol size indicates the size of the impactor.}
    \label{fig:xesc}
\end{figure}

As there is no exact solution in closed form nor a reasonably simple approximation, it must be determined numerically. After doing so for all model planets and impactor sizes of this study and for impact velocities from 5 to 60 km/s, we observed that for any given $g$ and $v_\mathrm{imp}$, the solutions $r_\mathrm{esc}$ of Eq.~\ref{eq:vef-r} are often only weakly dependent on the size of the impactor when normalized with it, i.e., $x_\mathrm{esc}=r_\mathrm{esc}/L$ does not change much with $L$ in many cases, especially at low $v_\mathrm{imp}$ and large $g$ (Fig.~\ref{fig:xesc}). With regard to the effect of the finite size of the impactor, it is more instructive to normalize $r_\mathrm{esc}$ with the radius $r_\mathrm{ej,min}$ within which material cannot be ejected as argued above, i.e., $x_\mathrm{esc}=r_\mathrm{esc}/r_\mathrm{ej,min}$. Then ejecta will only escape if $x_\mathrm{esc}>1$, i.e., if the corresponding point in Fig.~\ref{fig:xesc} sits above the dashed grey line. Furthermore, $x_\mathrm{esc}$ shows a quite clear dependence from $g$ and increases sublinearly with $v_\mathrm{imp}$. This suggests to construct a fit to the calculated values. We found that

\begin{equation}
x_\mathrm{esc}(g,v_\mathrm{imp})=[0.2293\exp(-0.1604g)+0.0927]\cdot v_\mathrm{imp}^{0.544},\label{eq:xesc-fit}
\end{equation}
where $g$ is given in m/s\textsuperscript{2} and $v_\mathrm{imp}$ is given in km/s, is a useful approximation to the solution if we neglect the influence of impactor size. It should be applicable at least to terrestrial planets with an internal structure as those considered here.
By solving the fitting formula for $v_\mathrm{imp}$ or $g$ at $x_\mathrm{esc}=1$, we find estimates for the lower bound on $v_\mathrm{imp}$ for a given $g$ and for the upper bound on $g$ for a given $v_\mathrm{imp}$ compatible with escaping ejecta, for all impactor sizes:

\begin{gather}
v_\mathrm{imp}\geq [0.2293\exp(-0.1604g)+0.0927]^{-1.839}\\
g\leq -6.2345\ln\left(4.3606v_\mathrm{imp}^{-0.544}-0.4042\right),
\end{gather}
with the condition $v_\mathrm{imp}<79.294$ km/s for the maximum $g$; for instance, only those of our principal models ($v_\mathrm{imp}=15$~km/s) for which $g\lesssim 3.234$~m/s\textsuperscript{2} will permanently lose ejecta and melt, i.e., only the smallest and lightest one ($R_\mathrm{P}=0.5R_\mathrm{E}, R_\mathrm{c}/R_\mathrm{P}=0.2$). For $v_\mathrm{imp}<8.032$ km/s
$g$ becomes negative, which may be interpreted as meaning that at such small impact velocities, no material could escape, no matter how small the target planet is. As the impact would cease to be a hypervelocity collision at much lower $v_\mathrm{imp}$, which was the assumption for Eq.~\ref{eq:vef} in the first place, this seems quite plausible.\par
With regard to the amount of melt that possibly escapes from the planet, we observe that for none of the cases we considered, $x_\mathrm{esc}$ exceeds 2.5, and that it is even less than 1.5 for almost all impacts with $v_\mathrm{imp}\leq 30$~km/s; in other words, $r_\mathrm{esc}\lesssim 0.9L$ except for very fast impactors. As material from inside $r_\mathrm{ej,min}=0.6L$ does not escape and the impactor is usually several times smaller than the transient crater, we can conclude that any escaping melt comes from a zone near the center of the crater that is only a small part of its total volume and that it is legitimate in general to neglect the escape of melt and other ejecta in mass balances.


\newpage

\begin{center}
\textbf{\Large Supporting information}
\end{center}
\vspace*{2ex}
\section{Melting efficiency scaling}
\begin{figure}[tb]
    \centering
    \includegraphics[width=\textwidth]{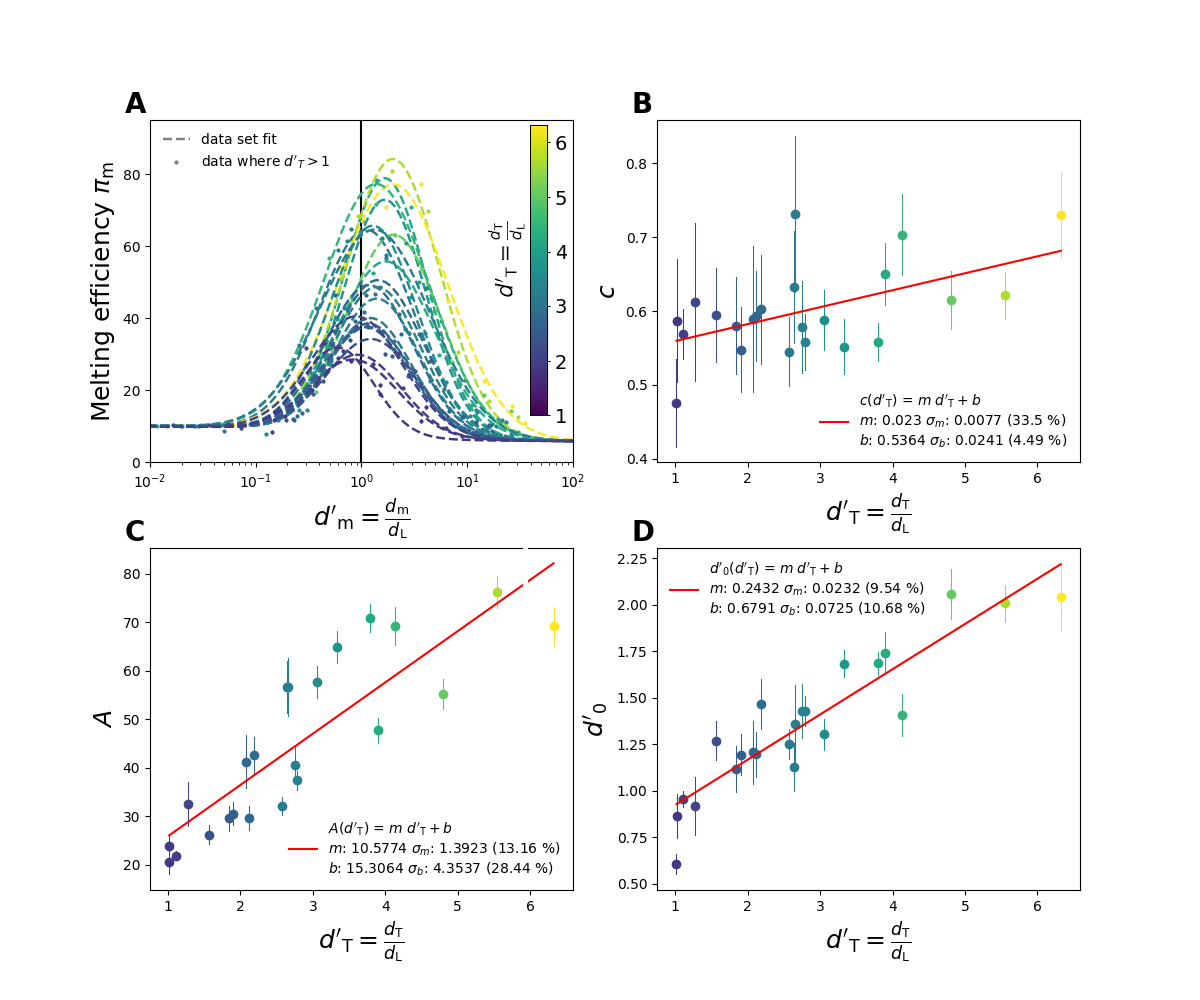}
    \caption{Individual scalings of the melting efficiency curves (A) and scaling parameters (B, C and D) used to calculate the empirical melting efficiency scaling law Eqs.~16--18.}
    \label{fig:fit-sup}
\end{figure}

In our approach to parameterizing our melting efficiency data, we fit each relevant melting efficiency curve ($d'_\mathrm{T} > 1$), consisting of multiple runs with varying impactor size for a given planet's thermal profile according to Eq.~16 as shown in panel A of Figure \ref{fig:fit-sup}. For each melting efficiency curve, the following parameters are fitted: $A$, giving the amplitude of the the bell-shaped peak (cf. Eq.~18) above the sigmoidal curve (cf. Eq.~17); $c$, describing the width of the bell-shaped curve; and $d'_0$, describing the offset of the melting depth $d_\mathrm{m}$ and hence of the centers of the peak and the sigmoidal function from the depth of the lithosphere $d_\mathrm{L}$. Each of these parameters for the individual data sets is shown along with its standard deviation in Figure \ref{fig:fit-sup} (panels B, C and D) as a function of $d'_\mathrm{T}$. Finally, we fitted the parameters by a linear regression as a function of $d'_\mathrm{T}$, resulting in the introduced scaling law. 

\end{document}